\documentclass[twocolumn]{aastex701}

\usepackage[utf8]{inputenc}
\usepackage{CJKutf8}
\newcommand\allthinstars{1515 }
\newcommand\allthickstars{378 }

\newcommand\allthinsingles{1121 }

\newcommand\allthinmultistars{394 }
\newcommand\allthinmultiplanets{862 }
\newcommand\allthickmultistars{98 }
\newcommand\allthickmultiplanets{207 }
\newcommand\allstars{1888 }
\newcommand\allplanets{2465 }
 
\begin{document}
\begin{CJK}{UTF8}{gbsn}

\title{The Orbital Eccentricities of Planets in the Kinematic Thin and Thick Galactic Disks}

\author[orcid=0000-0000-0000-0001,sname='Sagear']{Sheila Sagear}
\affiliation{Department of Astronomy, University of Florida, Gainesville, FL 32611}
\email[show]{ssagearastro@gmail.com}

\author[0000-0002-3247-5081]{Sarah Ballard}
\affiliation{Department of Astronomy, University of Florida, Gainesville, FL 32611}
\email{sarahballard@ufl.edu}

\author[0000-0003-2594-8052]{Kathryne J. Daniel}
\affiliation{Steward Observatory and Department of Astronomy, University of Arizona, 933 N. Cherry Avenue, Tucson, AZ 85721, USA}
\email{kjdaniel@arizona.edu}

\author[0000-0003-0872-7098]{Adrian M. Price-Whelan}
\affiliation{Center for Computational Astrophysics, Flatiron Institute, 162 5th Avenue, Manhattan, NY, USA}
\email{adrianmpw@gmail.com}

\author[0000-0002-6036-1858]{Sóley Ó. Hyman}
\affiliation{Steward Observatory and Department of Astronomy, University of Arizona, 933 N. Cherry Avenue, Tucson, AZ 85721, USA}
\email{soleyhyman@arizona.edu}

\author[0000-0003-0742-1660]{Gregory J. Gilbert}
\affiliation{Department of Astronomy, California Institute of Technology, Pasadena, CA 91125, USA}
\email{ggilbert@caltech.edu}

\author[0000-0002-1910-5641]{Christopher Lam}
\affiliation{Department of Astronomy, University of Florida, Gainesville, FL 32611}
\email{c.lam@ufl.edu}

\begin{abstract}

The orbital eccentricity distribution of exoplanets is shaped by a combination of dynamical processes, reflecting both formation conditions and long-term evolution. Probing the orbital dynamics of planets in the kinematic thin and thick Galactic disks provides insight into the degree to which stellar and Galactic environmental factors affect planet formation and evolution pathways. The classification of host stars in Galactic kinematic terms constitutes a potentially useful axis for the interpretation of orbital eccentricity, when included together with stellar metallicity and age. Leveraging the photoeccentric effect, we constrain orbital eccentricities for the sample of \textit{Kepler} planets and candidates orbiting F, G, K and M dwarf stars. With Gaia astrometry, inferred Galactic phase space information, and kinematic disk criteria calibrated on stellar chemical abundances, we probabilistically associate each planet host with the kinematic thin or thick Galactic disks. We then fit the underlying eccentricity distributions for the single- and multi-transit populations. We find that for single-transiting planets, kinematic thick disk planets exhibit higher eccentricities than thin disk planets, yet we find no such difference among multis. We determine that the difference in eccentricity is unlikely to be caused solely by the effects of host stellar metallicity or giant planet occurrence. We situate these findings in the context of known eccentricity relations, including its relationships with planet multiplicity, radius and metallicity. We suggest comprehensive analyses to disentangle these results from the effects of poorly understood star-planet relationships, such as that between stellar age and planetary orbital dynamics.

\end{abstract}

\section{Introduction}\label{sec:intro}

The \textit{Kepler} space telescope has to date enabled the discovery of over 4000 transiting planet candidates through long-term, precise photometric monitoring of a single field \citep{borucki_kepler_2010, koch_kepler_2010}, revealing the occurrence and characteristics of small planets in the Milky Way. The Gaia mission \citep{gaia_collaboration_gaia_2016, gaia_collaboration_gaia_2018} has also measured precise astrometric information for billions of stars, including the vast majority of known \textit{Kepler} planet hosts. Together, the \textit{Kepler} and Gaia missions produce a large sample of uniformly treated transit light curves along with homogeneously derived stellar parameters and astrometric information, uniquely suited for demographic analyses (see e.g. \citealt{berger_gaia-kepler_2020} and \citealt{fulton_california-kepler_2018}). 

The orbital dynamics of planets is of particular interest through a demographic lens, particularly the orbital eccentricity. Orbital eccentricity is a fundamental property of planetary systems: it quantifies a system's current dynamical state while serving as a signature of its dominant formation and evolution paths. Several large-scale orbital eccentricity trends have been discovered, such as the planet multiplicity—eccentricity relation: single-transit systems (hereafter ``singles") exhibit distinctly higher eccentricities than multi-transit systems (hereafter ``multis") \citep{limbach_exoplanet_2015, xie_exoplanet_2016, van_eylen_orbital_2019, sagear_orbital_2023}. Giant planets around metal-rich stars are also more likely to exhibit high eccentricities \citep{dawson_correlations_2016}, while there is some evidence that metal-poor stars more often host tightly-packed systems of small planets on near-circular orbits (so-called ``compact multis") at a higher rate than metal-rich stars \citep{anderson_higher_2021, brewer_compact_2018}. This relationship is also moderated by exoplanet radius and mass \citep{mills_california-kepler_2019}. There is also a strong relationship between orbital eccentricity and planet radius for stellar hosts across the FGKM spectral type range, such that planets with $R_p > 3.5 R_{\oplus}$ have markedly higher eccentricities than planets with $R_p < 3.5 R_{\oplus}$ \citep{gilbert_planets_2025, sagear_orbital_2025}. 

Large transit survey data combined with \textit{Gaia} stellar and astrometric information have revealed demographic trends between planetary systems and their host stars, but the star-planet relationship is only part of the story. Recent studies have begun placing the relationship between metallicity and planet demographics in the context of the host stars' Galactic position and dynamical history \citep{winter_stellar_2020, kruijssen_bridging_2020, nielsen_planet_2023, kruijssen_not_2021, bashi_exoplanets_2022, yang_planets_2023}. The Galactic context has assumed increasing importance in the full story of planet formation and evolution, even potentially late-stage dynamical evolution. Emerging evidence suggests that planetary systems retain some information of their place in the Galaxy: for example, \citet{zink_scaling_2023} determined that the occurrence rate of Kepler and K2 planets decreases with increasing Galactic oscillation amplitude in ways not attributable to metallicity alone. Stellar flybys and other external perturbations may affect the dynamical evolution of systems: there is theoretical support for such flybys to either directly impact outer planets and potentially propagate inward \citep{zakamska_excitation_2004, rodet_correlation_2021, veras_exoplanets_2013, schoettler_effect_2024, charalambous_breaking_2025, mctier_8_2020}. Alternatively, flybys and Galactic tides will both play a role in exoplanets in wide binary systems, whose altered orbit might then disrupt planets \citep{kaib_planetary_2013, correa-otto_galactic_2017}.

Leveraging the \textit{Kepler} sample of transiting planets and candidates along with Gaia astrometric information, we investigate the orbital eccentricities of exoplanets as a function of Galactic dynamical properties: specifically, membership in the kinematic thin vs. thick disks. In this work, we investigate the extent to which planetary orbital eccentricities for thin and thick disk host stars appear to be drawn from distinct parent distributions. Such differences would be key to understanding whether distinct formation and evolution processes dominate for kinematic thin and thick disk systems, and informs future studies on the degree of influence stellar age, close stellar encounters, or birth environment has on the dynamical state of exoplanets. 

This manuscript is organized as follows. In Section \ref{sec:sampleselection}, we introduce the stellar and planetary sample selection process for this analysis, along with the criteria for Galactic disk classification. In Section \ref{sec:methods}, we describe the methods and tools we use to fit the sample of \textit{Kepler} transit light curves and the statistical methods we use to analysis the population-level underlying eccentricity distributions. We present and discuss implications of our results in Sections \ref{sec:results} and \ref{sec:discussion}, and conclude in Section \ref{sec:conclusion}.

\section{Sample Selection} \label{sec:sampleselection}

\begin{figure*}
    \centering
    \includegraphics[width=\linewidth]{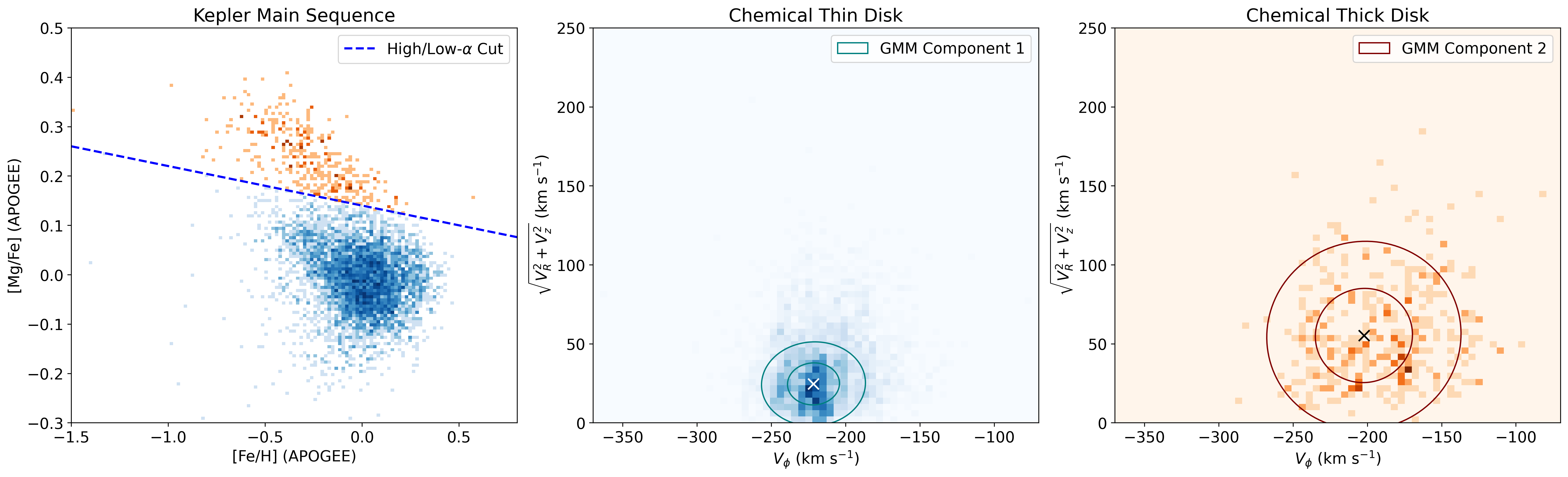}
    \caption{\textit{Left:} Magnesium [Mg/Fe] vs. iron [Fe/H] abundances from ASPCAP for the calibration sample of Kepler/APOGEE main-sequence stars. The high-$\alpha$/low-$\alpha$ linear cutoff is shown with the blue dotted line. The green heatmap represents the low-$\alpha$ (thin disk) calibration sample, and the red heatmap represents the high-$\alpha$ (thick disk) calibration sample. \textit{Center:} The thin disk calibration sample in $V_{\phi}$ vs. $\sqrt{V_{r}^2 + V_{z}^2}$ velocity space. The mean of the low-$\alpha$ component of the GMM is designated with a cross, and the concentric circles represent the $1-$ and $2-\sigma$ regions. \textit{Right:} Same as \textit{Center}, but for the high-$\alpha$ calibration sample.}
    \label{fig:chemical}
\end{figure*}

\begin{figure}
    \centering
    \includegraphics[width=\linewidth]{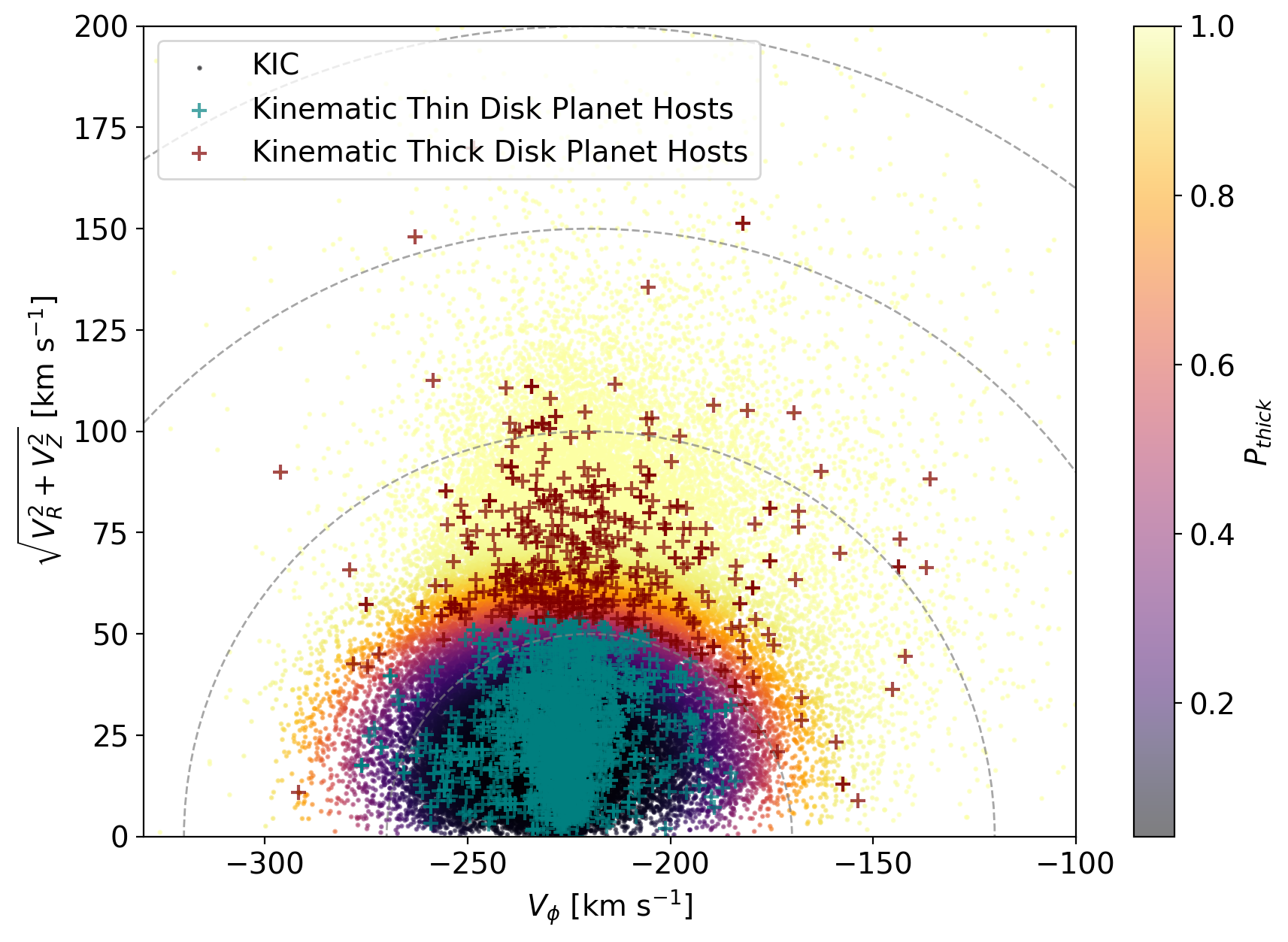}
    \caption{Toomre diagram of KIC field stars (light blue) and the Kepler planet host sample used in this work. The color represents the kinematic thick disk association probability $P_\mathrm{thick}$ for each star. Thin disk stars are designated with teal crosses and thick disk stars are designated with red crosses. The gray dotted lines represent lines of constant velocity at $V_{tot} = 50,100,150,$ and $200$ $[\mathrm{km}]$ $\mathrm{s}^{-1}$.}
    \label{fig:toomre}
\end{figure}

Beginning with the entire sample of Kepler Objects of Interest (KOIs), we first remove any planets marked as False Positives according to the NASA Exoplanet Archive \citep{christiansen_nasa_2025}. We select planets with orbital periods between 1 and 100 days. To eliminate stars with a high likelihood of being unresolved stellar binaries, we filter out stars with greater than $5\%$ flux contamination from nearby sources as reported by \citet{furlan_lessigreaterkeplerlessigreaterfollow-up_2017} or with a Gaia renormalized unit weight error (RUWE) value of $> 1.4$. The kinematic disk classification method is described in Section \ref{subsec:thinandthickdiskprobabilities}. Our sample consists of \allplanets total planets or planet candidates orbiting \allstars stars. We have \allthickstars kinematic thick disk single hosts, \allthickmultiplanets kinematic thick disk multi planets orbiting \allthickmultistars stars, \allthinsingles kinematic thin disk single hosts, and \allthinmultiplanets kinematic thin disk multi planets orbiting \allthinmultistars stars.

\begin{figure}
    \centering
    \includegraphics[width=\linewidth]{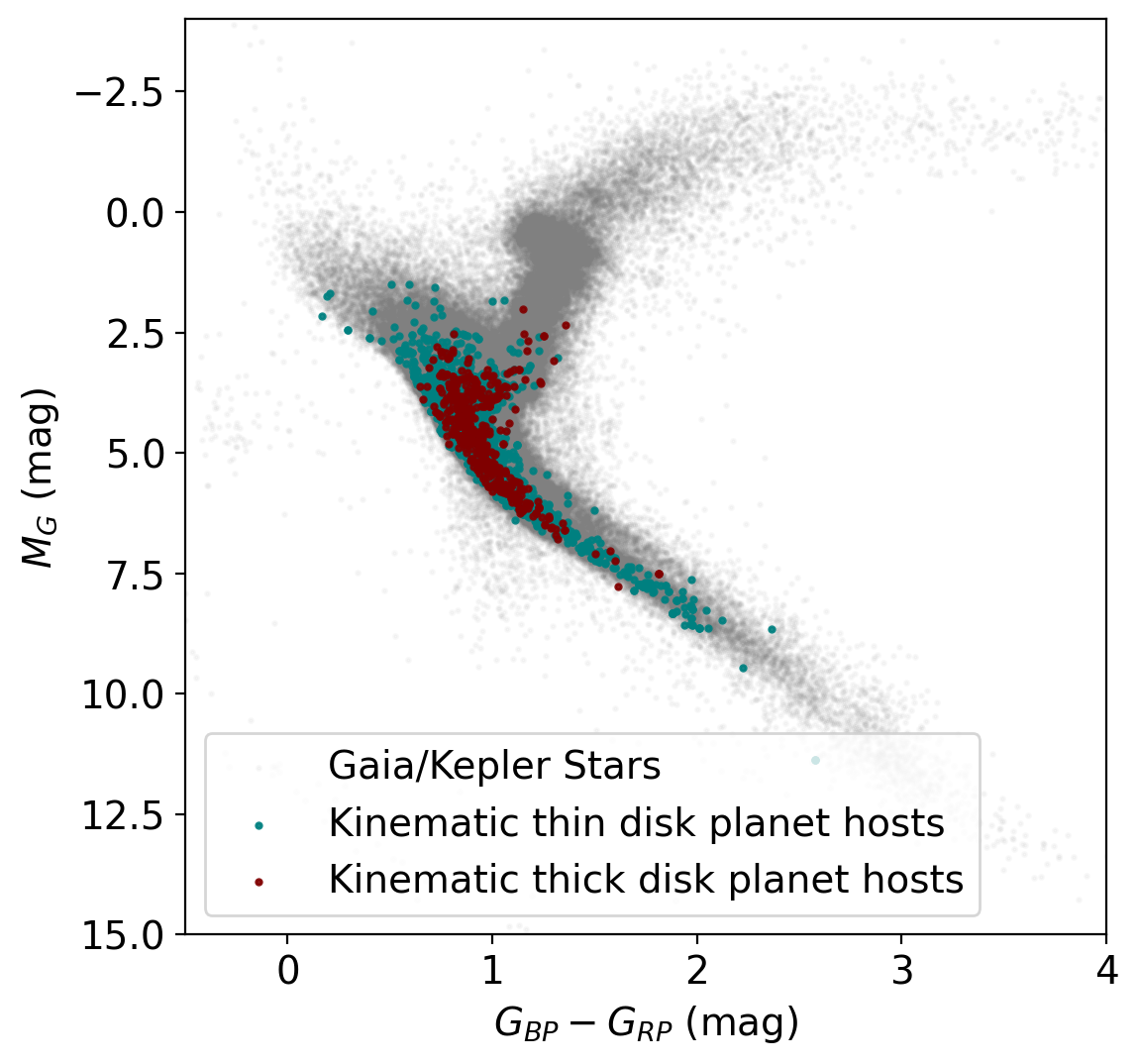}
    \caption{Color-magnitude diagram of the KIC sample (gray), the sample of thin disk planet hosts (teal), and the sample of thick disk planet hosts (red). The absolute magnitude $M_G$ and color $G_{BP}-G_{RP}$ are calculated with Gaia DR3 data.}
    \label{fig:colormag}
\end{figure}

\begin{figure*}[ht!]
    \centering
    \includegraphics[width=\linewidth]{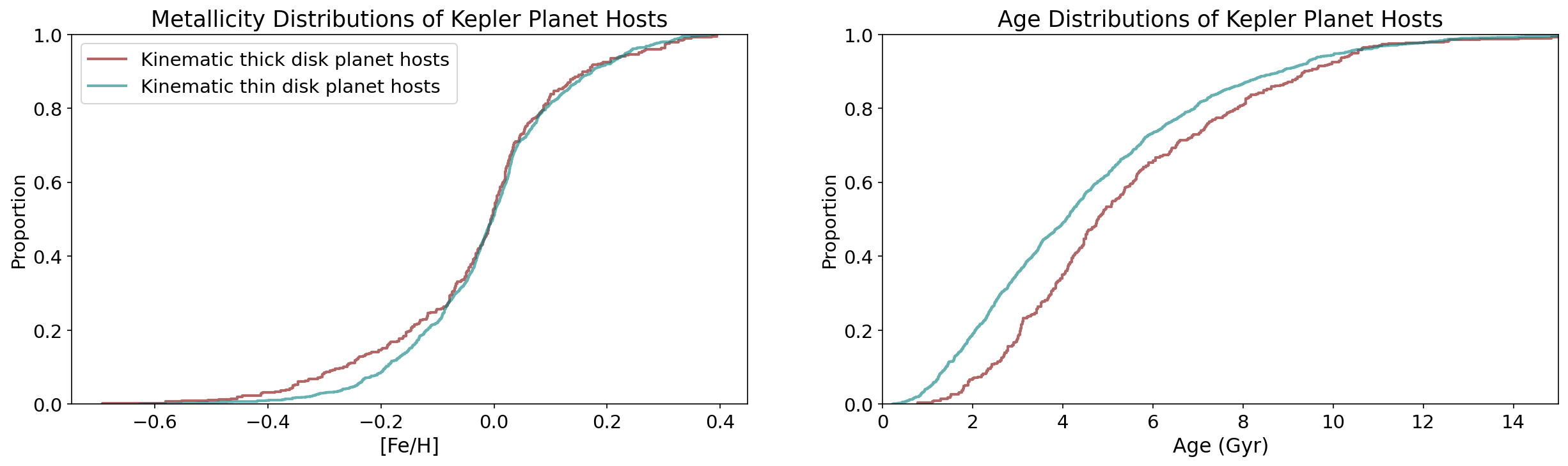}
    \caption{\textit{Left:} CDFs of stellar metallicity $[Fe/H]$ from from \citet{berger_gaia-kepler_2020}. The sample of thick disk planet hosts is shown in red and thin disk planet hosts are shown in teal. \textit{Right:} CDFs of stellar age (in Gyr, from \citealt{berger_gaia-kepler_2020}) for thin and thick disk planet hosts with $T_\mathrm{eff}>4700K$. }
    \label{fig:thick_thin_age_metal}
\end{figure*}

We present the distributions of various stellar sample parameters for the entire kinematic thin and thick disk samples, as well as for the subsamples of single and multi hosts. In Figure \ref{fig:colormag}, we show the color-magnitude diagram of the kinematic thin and thick disk planet host samples constructed with Gaia DR3 data. In Figure \ref{fig:thick_thin_age_metal}, we show the CDFs of stellar iron abundance [Fe/H] from \citealt{berger_gaia-kepler_2020} for each sample. The entire sample is centered close to solar metallicity ($[\mathrm{Fe/H}] = 0$). We also show the distributions of stellar ages as determined by \citet{berger_gaia-kepler_2020} for those stars with $T_\mathrm{eff} > 4700$ K. It is clear that the kinematic thick disk sample is generally composed of older stars than the thin disk sample, as we expect. 

Stellar populations vary throughout the Galaxy in relation to Galactic dynamics, with young and metal-rich stars concentrated in the thin Galactic disk and spiral arms and older, metal-poor populations associated with higher vertical dynamical properties. Planetary dynamical heat is correlated with host stellar metallicity and likely also stellar age, though the latter relationship is not yet well-defined. However, the extent to which individual stellar properties (namely, stellar metallicity and age) influence planetary dynamics independently from Galactic dynamics remains unknown. The sample of planet hosts in this work is largely limited to main-sequence stars; however, stellar age and metallicity may affect planetary orbits independently from Galactic dynamics, even within the main-sequence. In this work, we do not fully disentangle the influence of stellar properties and Galactic dynamics from the orbital eccentricities of planets; defining these more subtle effects an investigation will require careful statistical treatment of the data and will be explored in future work.

\subsection{Thin and Thick Disk Probabilities} \label{subsec:thinandthickdiskprobabilities}

We determine thick vs. thin disk associations for planet hosts in our sample using kinematic disk probabilities calibrated on chemical disk associations. To do this, we must analyze the full 6-D Galactic phase-space information of stars (3-D positional information with 3-D velocity information). From Gaia, precise positions, parallaxes, and proper motions for virtually all nearby stars are readily available; however, radial velocity (RV) measurements are only available for about half of planet-hosting stars in our sample. \citet{angus_3d_2022} present a method to infer 3-D stellar velocities by marginalizing over missing RVs. The authors use a prior constructed from the stellar velocity distributions of the portion of the Kepler sample with available RVs from Gaia; therefore, the values presented in \citet{angus_3d_2022} are constructed for Kepler stars. We use the 3-D velocities presented by \citet{angus_3d_2022} to calculate thin and thick disk probabilities for each planet host in our sample. Where Gaia RV data is available, we use the 3-D velocities calculated directly from the Gaia 6-D phase space coordinates $(\alpha_{0}, \delta_{0}, \varpi_{0}, \mu_{\alpha0}, \mu_{\delta 0}, \mu_{r0})$. Only where Gaia RV data is unavailable do we utilize the inferred 3-D velocities.

We convert the Galactocentric Cartesian velocities presented in \citet{angus_3d_2022} to Galactocentric cylindrical coordinates. \citet{angus_3d_2022} present a comprehensive validation that the inferred 3-D velocities are consistent with the Gaia 3-D velocities for stars in the Kepler field. We use the cylindrical $(V_\phi, V_r,V_z)$ velocities to calculate the relative likelihood of each star to occupy the thin or thick disk populations. Several recent studies, such as those by \citet{chen_planets_2021} and \citet{bashi_exoplanets_2022}, recalibrate the kinematic disk classification methods presented by \citet{bensby_elemental_2003, bensby_exploring_2014} for various bins of Galactic cylindrical positions $R$ and $Z$ in an effort to apply these disk classifications to the Kepler field. We wish to apply a classification method robust to the entire KIC (of which over 40\% have a distance of over 1 kpc) and future surveys, such as the Nancy Grace Roman Galactic Exoplanet Survey \citep{johnson_predictions_2020, wilson_transiting_2023} and Gaia Data Release (DR) 4, which will include significant samples of stars far from the Solar neighborhood. Therefore, we elect not to use Local Standard of Rest (LSR) stellar velocities (as \citealt{bensby_exploring_2014, chen_planets_2021}; and \citealt{bashi_exoplanets_2022}), since the LSR is a poor approximation for stellar velocities far from the Solar neighborhood, and re-calibration of LSR disk classification criteria using cylindrical positions potentially introduces additional systematic biases. Instead, we elect to use only Galactocentric cylindrical velocities to describe stellar motion, where the rest frame is the Galactic center and stellar velocities are agnostic to Solar Galactic motion. 

In order to robustly classify stars with the ``thin" and ``thick" disk populations, chemical abundances along with Galactic kinematic information are typically used in tandem. However, since many Kepler stars do not have reliable or homogeneously derived chemical abundances, we use available chemical abundances to inform kinematic criteria which we may apply to stars without spectroscopic data. We generate a calibration sample by first crossmatching the Kepler sample of planet hosts with data from the Apache Point Observatory Galactic Evolution Experiment (APOGEE) DR17 \citep{gunn_25_2006, zasowski_target_2013, nidever_data_2015, majewski_apache_2017, zasowski_target_2017, wilson_apache_2019, beaton_final_2021, santana_final_2021, abdurrouf_seventeenth_2022}. 
We take stellar chemical abundances, specifically the [Mg/Fe] and [Fe/H] abundances, from the APOGEE Stellar Parameters and Abundances Pipeline (ASPCAP) \citep{shetrone_sdss-iii_2015, garcia_perez_aspcap_2016, smith_apogee_2021}. We apply stellar parameter cuts to ensure only main-sequence stars are included in the calibration sample ($T_\mathrm{eff} < 6500$, $\mathrm{log}(g) > 4.0$). We then apply a linear high/low-$\alpha$ cutoff point to demarcate the chemical thick and thin disk samples, defined by the line $\mathrm{[Mg/Fe]} = -0.08*\mathrm{[Fe/H]} +0 .14$ (Figure \ref{fig:chemical}, left). We plot the resulting high-$\alpha$ (chemical thick disk) and low-$\alpha$ (chemical thin disk) samples in $(V_\phi, \sqrt{V_r^2+V_z^2})$ space and fit a two-component Gaussian Mixture Model (GMM) to the sample. The chemical thin disk and chemical thick disk subsamples, along with their respective components of the GMM, are shown individually in Figure \ref{fig:chemical} (center and right). We apply the trained GMM to the entire Kepler sample (with velocities calculated from Gaia DR3 where available, otherwise with inferred velocities from \citealt{angus_3d_2022}) based on each star's Galactocentric cylindrical velocities. Thus, every KIC star has an associated thick disk probability $(P_\mathrm{thick},$ where $P_\mathrm{thin}=1-P_\mathrm{thick})$. This thick disk probability is based purely on stellar kinematic cuts, which have been calibrated using stellar chemical abundances.
We take ``thick disk" stars to be those where $P_{thick} > 0.5$ (that is, those that are more likely to be kinematically associated with the high-$\alpha$, chemical thick disk sample than the low-$\alpha$, chemical thin disk sample) and ``thin disk" stars to be those where $P_{thick} < 0.5$. In total, we have in our sample \allthinstars thin disk planet hosts and \allthickstars thick disk planet hosts. We present a Toomre diagram of the KIC sample along with the kinematic thin disk and thick disk planet hosts in Figure \ref{fig:toomre}.

\section{Methods} \label{sec:methods}

\subsection{Constraining Eccentricities from Transit Light Curves} \label{subsec:thephotoeccentriceffect}

In order to generate eccentricity constraints for our sample of Kepler planets, we leverage the photoeccentric effect. The photoeccentric effect was first posited by \citet{ford_characterizing_2008, kipping_novel_2012}, and \citet{dawson_photoeccentric_2012} and has since been leveraged by a number of demographic analyses of orbital eccentricities \citep[e.g.]{van_eylen_orbital_2019, sagear_orbital_2023, gilbert_planets_2025, sagear_orbital_2025}. It describes a phenomenon by which a planet's orbital eccentricity can be inferred by precise measurement of its transit duration combined with a stellar density prior. The photoeccentric formalism is fully described in \citet{dawson_photoeccentric_2012}. 

The main benefit to using the photoeccentric method is that we generate eccentricity posteriors for virtually all known transiting planets, resulting in a larger sample than would be possible using, for example, only eccentricities constrained by radial velocity. Sample size is critical within any demographic analysis, but especially when investigating exoplanets in the Galactic context, as there are relatively few known planets outside the thin disk. However, the main drawback is that the eccentricity posteriors generated from transit light curves are necessarily poorly constrained, due to limited light curve precision, knowledge of the stellar density, and the degeneracy between eccentricity and longitude of periastron ($\omega$) (the transiting system's viewing angle). Therefore, investigating eccentricities by the photoeccentric effect is most useful when probing the underlying eccentricity distribution for a population of planets within a hierarchical Bayesian framework rather than measuring eccentricities for individual planets.

We constrain orbital eccentricities for our planetary sample using Kepler transit light curves along with a host stellar density prior. Care must be taken in selecting a stellar density prior, as the accuracy and precision of the resulting eccentricity posteriors rely on the accuracy and precision of the stellar density. Prioritizing homogeneity, we take the stellar sample and stellar densities presented by \citet{berger_revised_2018}.

We detrend and fit the Kepler transit light curves in a homogeneous manner using the \texttt{ALDERAAN} transit fitting package. We briefly describe the \texttt{ALDERAAN} pipeline here; further details on the pipeline can be found in \citet{gilbert_planets_2025}. The primary benefits of using the \texttt{ALDERAAN} pipeline are uniform treatment of all light curves and transit timing variations (TTVs), the use of nested sampling \citep{skilling_nested_2004, skilling_nested_2006, higson_dynamic_2019} through \texttt{dynesty} \citep{speagle_dynesty_2020, koposov_joshspeagledynesty_2024} rather than Markov Chain Monte Carlo (MCMC) inference, and improved computational efficiency compared to gradient-descent transit-fitting algorithms such as \texttt{exoplanet} \citep{foreman-mackey_exoplanet_2021} and \texttt{jaxoplanet} \citep{hattori_exoplanet-devjaxoplanet_2024}. It has been shown by \citet{gilbert_implicit_2022} that the use of gradient descent (MCMC) transit fitting algorithms may bias the resulting transit posteriors, particularly the impact parameter: the use of nested sampling mitigates these biases by fully exploring high-dimensional posteriors with complex shapes.

We download the Kepler PDCSAP light curves from MAST and stitch individual Kepler quarters together, normalize them, and remove invalid flux values. Modeling transit timing variations (TTVs) is critical in estimating eccentricities, as improperly accounting for TTVs will result in a biased transit duration measurement which directly affects the eccentricity estimate. The \texttt{ALDERAAN} pipeline implements an iterative template-matching approach (as in \citealt{mazeh_transit_2013} and \citealt{holczer_transit_2016}) by which the transit shape parameters are first estimated using a fixed ephemeris, then the individual transit times are subsequently varied with the shape parameters fixed. We fit each planet with a set of ephemeris free parameters and transit shape free parameters. The ephemeris free parameters are $\mathrm{C}0$ and $\mathrm{C}1$ (linear perturbations to the transit timing model). The shape free parameters include the planet—star radius ratio $R_p/R_s$, the impact parameter $b$, and full transit duration from first to fourth contact $T_{14}$ in addition to the system-level quadratic limb darkening parameters {$u_1, u_2$}. The entire set of free parameters \{$\mathrm{C}0, \mathrm{C}1, R_p/R_s,b,T_{14}$\} were sampled simultaneously for all planets in every system, ensuring system-wide consistency in each posterior point. This is especially important for the limb-darkening free parameters, which should be consistent across all planets orbiting the same star. The transit fit free parameters and their priors are listed in Table \ref{table:transitfitfreeparameters}.

We fit the processed Kepler transit light curves using nested sampling implemented with \texttt{dynesty} \citep{speagle_dynesty_2020} within the \texttt{ALDERAAN} package. We ensure that the light curve models are integrated appropriately over the Kepler long- and short-cadence exposure times. We visually inspect each transit fit for quality or convergence issues, and we remove less than $2 \%$ of the sample due to non-convergent posteriors. To extract $e$ and $\omega$ constraints from the transit model posteriors, we implement the post-model importance sampling technique introduced by \citet{macdougall_accurate_2023} and previously implemented by \citet{gilbert_planets_2025} and \citet{sagear_orbital_2025}. The full formalism for the importance sampling method can be found in \citet{macdougall_accurate_2023} and further validation testing found in \citet{sagear_orbital_2025}. 




\begin{deluxetable*}{ccc}
\tablecaption{Transit fit free parameters, prior distributions and values. \label{table:transitfitfreeparameters}}
\tablehead{\colhead{Free Parameter} & \colhead{Description} & \colhead{Prior}}
\startdata
$C_0, C_1$ & Linear perturbations to the transit timing model & $\sim \mathcal{N} (0, 1)$ \\ 
$\mathrm{T_{14}}$ & First-to-fourth contact transit duration & $\sim \log\mathcal{U}^\dagger$ \\
$R_p/R_s$ & Planet-star radius ratio & $\sim \mathcal{U} (10^{-5}, 0.99)$ \\
$b$ & Impact parameter & $\sim \mathcal{U}(0, 1 + R_p/R_\star)$ \\
$(u_1, u_2)$ & Quadratic limb darkening parameters & $\sim \mathcal{N} (\mu, 0.1)$ \\
\enddata
\tablecomments{Transit duration $T_{14}$ was sampled between one Kepler short cadence integration length (1 minute) and three times the transit duration found during TTV fitting (i.e. the size of the nominal transit window). Limb darkening coefficient priors are determined using values derived from Gaia measurements \citep{gaia_collaboration_gaia_2018, berger_gaia-kepler_2020} and stellar atmosphere models \citep{husser_new_2013, parviainen_ldtk_2015}.}
\end{deluxetable*}

\subsection{Hierarchical Bayesian Inference} \label{subsec:hierarchicalbayesianinference}

The decision of model choice within hierarchical Bayesian inference remains a difficult one, especially when a distribution has no known functional form in nature. This is the case for the underlying orbital eccentricity distribution. There is a strong physical motivation for the use of the Rayleigh distribution: as posited in 
\citep{sagear_orbital_2025}, if the orbital mid-planes of planetary systems are distributed normally with respect to our viewing angle, and individual inclinations are distributed normally around the orbital mid-plane, then mutual inclinations must be distributed according to a Rayleigh distribution. If orbital eccentricities follow the distribution of mutual inclinations according to a system's maximum angular momentum deficit (AMD) stability limit \citep{he_architectures_2020, millholland_evidence_2021}, then orbital eccentricities must be distributed in a similar pattern to mutual inclinations -- thus, they would also be distributed according to a Rayleigh distribution. However, as further discussed in \citet{sagear_orbital_2025} along with other demographic analyses of orbital eccentricities \citep[e.g.]{van_eylen_orbital_2019, gilbert_planets_2025}, the underlying eccentricity distribution does not consistently appear to resemble a Rayleigh distribution to first order. This could be due to a number of factors including the orbital eccentricity distribution resembling another shape entirely, but could also be caused by the lack of planets with precise eccentricity constraints we have in hand today (which are only for 1000 to 2000 planets). In \citet{sagear_orbital_2025}, the authors acknowledge the benefits of using an empirical model, or a model which bears strong resemblance to the empirical model. We suggest that the most prudent treatment of the data is to fit and analyze several different model choices using the same posterior data, as well as to release the full posterior information for further independent analysis. We assess the quality of each model fit with the Bayesian Information Criterion (BIC) in Section \ref{subsec:BIC}.

With individual eccentricity posteriors in hand and a parent distribution model of choice, we infer the underlying population-level eccentricity distribution within a Bayesian hierarchical framework. This method has been used previously to infer population-level eccentricity distributions from individually unconstraining eccentricity posteriors by \citet[e.g.]{van_eylen_orbital_2019, sagear_orbital_2023, gilbert_planets_2025}, and \citet{sagear_orbital_2025}. Details of the Bayesian hierarchical modeling techniques used here can be found in \citet{gilbert_planets_2025} and \citet{sagear_orbital_2025} and references therein. We note that it is critical to account for the non-uniform geometric transit probability as a function of eccentricity: highly eccentric planets have a higher transit probability than their counterparts on circular orbits \citep{barnes_ages_2007, burke_impact_2008, kipping_bayesian_2014}. We properly account for the non-uniform transit probability across $e$ and $\omega$ within our Bayesian framework as in \citet{sagear_orbital_2023}, \citet{gilbert_planets_2025} and \citet{sagear_orbital_2025}.




We alternately assume a Rayleigh distribution, half-Gaussian, Beta distribution, and monotonic Beta distribution with modified priors for the underlying eccentricity distribution. The Rayleigh distribution depends on one parameter, $\sigma_R \sim U(0,1]$. The half-Gaussian distribution is a normal distribution with mean fixed at 0 and has one free parameter, $\sigma_{HG} \sim U(0,1]$. The Beta and monotonic Beta distribution each rely on two free parameters $\alpha$ and $\beta$. The monotonic Beta distribution was used in \citet{sagear_orbital_2025} as the primary model choice for population eccentricity distributions and is defined in the range $[0, 1]$ with a modified prior which mandates the function always peaks at 0 and monotonically decreases ($\alpha < 1$ and $\beta > 1$). Improper choices for prior distributions when sampling across Bayesian hierarchical models can have severe effects on the resulting distributions, especially in modeling underlying eccentricity distributions as Beta distributions \citep{nagpal_impact_2023, do_o_orbital_2023}. As in \citet{sagear_orbital_2025}, we apply the parameterization $\tau = 1/\sqrt{\alpha + \beta}$ and $\mu = \alpha/(\alpha + \beta)$. For the Beta distribution, we simply apply the priors $\tau \sim U(0,1]$ and $\mu \sim U(0,1]$. For the monotonic Beta distribution, we enforce $\alpha < 1$ and $\beta > 1$ by applying the priors $\tau \sim U(0,1]$ and $\mu \sim U(0,\mathrm{min}(\tau^2, 1-\tau^2)]$. For both the Beta and monotonic Beta distributions, we sample with an off-centered parameterization to reduce the effects of strong correlations between parameters \citep{betancourt_hamiltonian_2013}. We sample directly from $\mu_{\textrm{\small latent}}$ with the prior $\mu_{\textrm{\small latent}} \sim U(0,1]$, where $\mu = \mu_{l} + \mu_{\textrm{\small latent}}(\mu_{l}+\mu_{u})$ and $\mu_{l}$ and $\mu_{u}$ are respectively the lower and upper bounds of the $\mu$ prior.

For the subsamples of singles and multis within the kinematic thin and thick disk, we generate the best-fit underlying eccentricity distributions from each planet's eccentricity posterior samples. We perform the population distribution modeling with gradient descent using the \texttt{numpyro} Python package \citep{phan_composable_2019} using two chains with 1000 steps each. We check for convergence using the Gelman-Rubin statistic $\hat{R}$ \citep{gelman_inference_1992} and ensure that for all free parameters $\hat{R}<1.05$.

\subsection{The Bayesian Information Criterion} \label{subsec:BIC}

We quantify the quality of the Rayleigh, Beta, monotonic Beta and Half-Gaussian model fits by analyzing the Bayesian Information Criterion (BIC) \citep{schwarz_estimating_1978}. Overall, we find that the Rayleigh distribution provides the best fit to the data compared to all four models for both the kinematic thin and thick disks. For both kinematic disk samples, we find that the Beta and monotonic Beta distributions provide fits of nearly equivalent quality to one another ($\Delta BIC < 1$), but compared to the Rayleigh distribution are strongly disfavored ($\Delta BIC \sim 20$). The half-Gaussian distribution is also strongly disfavored compared to the Rayleigh distribution ($\Delta BIC > 25$). 

\subsection{Leave \textit{N} Out Validation} \label{subsec:leavenoutvalidation}

Within this hierarchical Bayesian inference technique, small numbers of erroneously high eccentricity posteriors could significantly affect the resulting underlying distribution, and individual eccentricity posteriors are themselves susceptible to inflation. We visually inspect each transit fit by hand. To ensure that our results are robust to this effect, we also perform a leave \textit{N} out validation test. For each fit presented in this work, we remove 10 \% of the sample and re-fit the hierarchical eccentricity models. We do this 10 times for each sample. We ensure that all resulting best-fit distributions are consistent with one another. The expected value $\langle e \rangle$ is affected by less than $5\%$ with each ``leave \textit{N} out" iteration.

\section{Results} \label{sec:results}

\begin{figure*}[ht!]
    \centering
    \includegraphics[width=\linewidth]{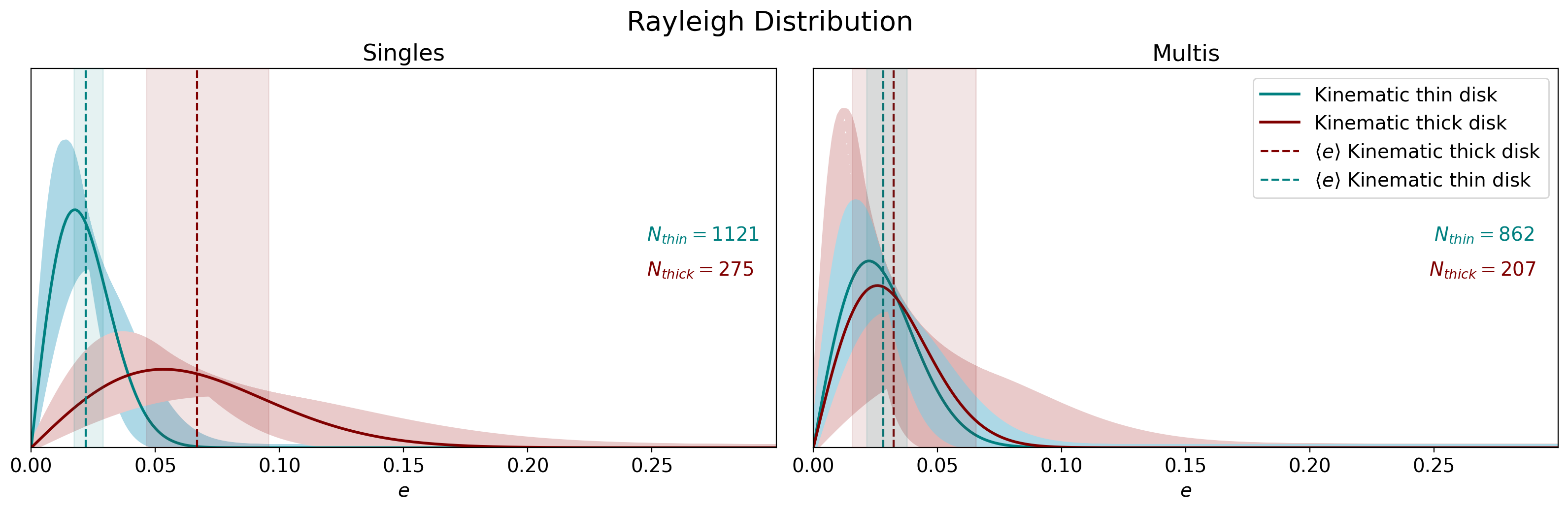}
    \caption{Underlying eccentricity distribution of kinematic thick vs. thin disk stars, modeled as Rayleigh distributions.  The thick disk $e$ distributions are shown in red, and the thin disk $e$ distributions are shown in blue. The shaded regions represent the $1\sigma$ uncertainties. The dashed lines represent the location of the expected value $\langle e \rangle$ for each distribution, and the shaded vertical regions correspond to the $1\sigma$ uncertainty on $\langle e \rangle$.}
    \label{fig:rayleigh}
\end{figure*}

\begin{figure*}[ht!]
    \centering
    \includegraphics[width=\linewidth]{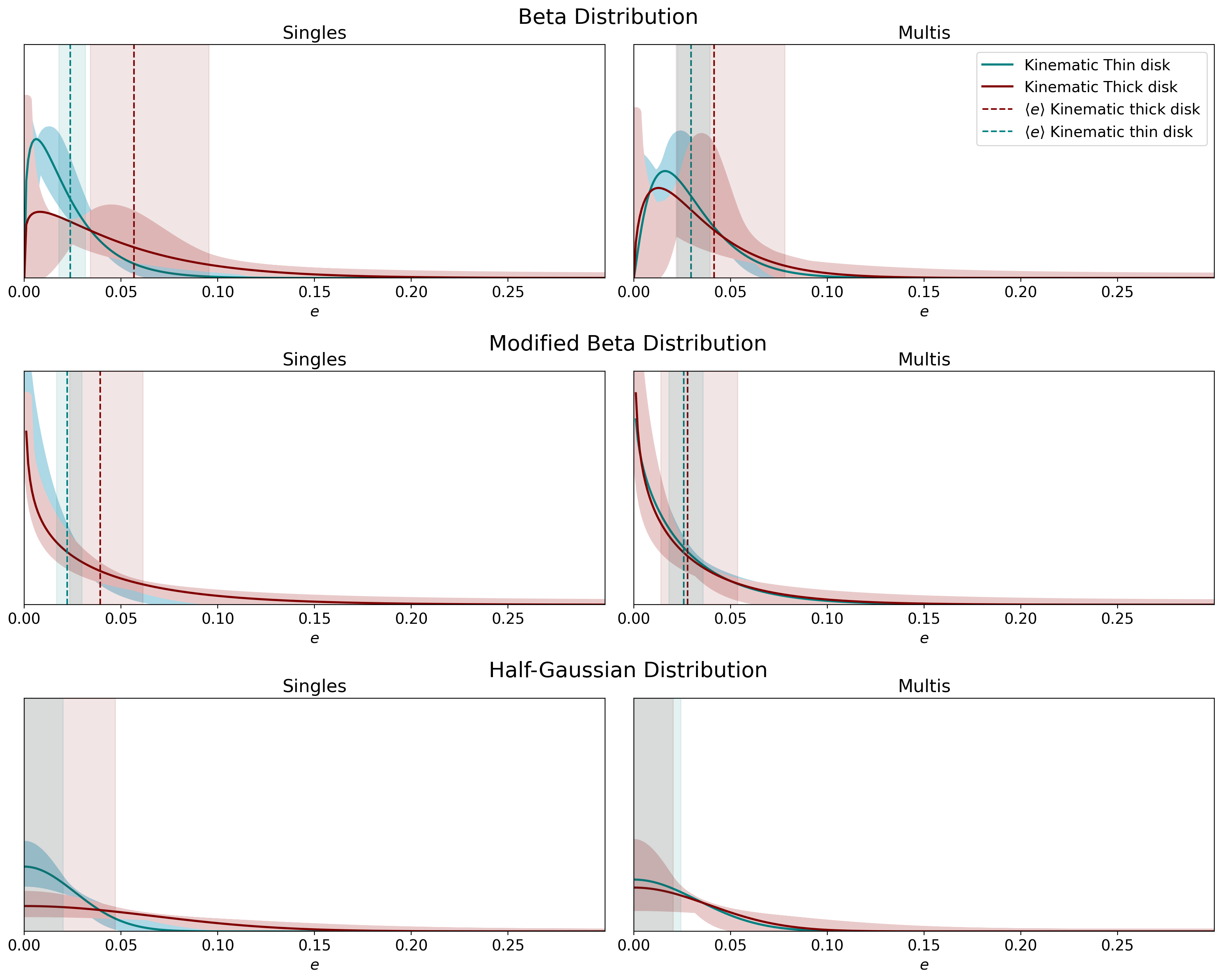}
    \caption{Underlying eccentricity distribution of kinematic thick vs. thin disk stars, modeled as Beta, modified Beta, and half-Gaussian distributions, for singles and multis. The thick disk $e$ distributions are shown in red, and the thin disk $e$ distributions are shown in blue. The shaded regions represent the $1\sigma$ uncertainties. The dashed lines represent the location of $\langle e \rangle$ for each distribution, and the shaded vertical regions correspond to the $1\sigma$ uncertainty on $\langle e \rangle$.}
    \label{fig:allotherdists}
\end{figure*}

We present the underlying eccentricity distributions for planets in the kinematic thin disk vs. planets in the kinematic thick disk. In Figure \ref{fig:rayleigh}, we present the underlying eccentricity distributions modeled as Rayleigh distributions for singles and multis within the kinematic thick and thin disk samples. In Figure \ref{fig:allotherdists}, we present these underlying eccentricity distributions modeled as Beta distributions with priors $\alpha \sim U(0,1), \beta \sim U(0,1)$, and monotonic Beta distributions with priors $\alpha \sim U(0,1), \beta \sim U(1, \infty)$, and half-Gaussian distributions. In Table \ref{table:modelfits}, we present the best-fit parameters and uncertainties for each model.

\begin{deluxetable*}{cccc}
\tablecaption{Best-fit underlying $e$ distribution parameters \label{table:modelfits}}
\tablehead{\colhead{Model} & \colhead{Kinematic Disk Association} & \colhead{Transit Multiplicity} & \colhead{Expected Value $\langle e \rangle$}}
\startdata
Rayleigh & Thick & Singles & $0.066_{-0.021}^{+0.030}$ \\ 
\nodata & Thick & Multis & $0.033_{-0.018}^{+0.032}$ \\
\nodata & Thin & Singles & $0.022_{-0.005}^{+0.007}$ \\ 
\nodata & Thin & Multis & $0.030_{-0.007}^{+0.001}$ \\ 
\hline
Beta & Thick & Singles & $0.058_{-0.023}^{+0.039}$\\
\nodata & Thick & Multis & $0.042_{-0.022}^{+0.035}$ \\ 
\nodata & Thin & Singles & $0.023_{-0.006}^{+0.008}$ \\ 
\nodata & Thin & Multis & $0.030_{-0.008}^{+0.010}$ \\  
\hline
Monotonic Beta & Thick & Singles & $0.041_{-0.016}^{+0.025}$ \\ 
\nodata & Thick & Multis & $0.025_{-0.012}^{+0.023}$ \\ 
\nodata & Thin & Singles & $0.022_{-0.006}^{+0.008}$ \\ 
\nodata & Thin & Multis & $0.026_{-0.008}^{+0.010}$ \\ 
\hline
Half-Gaussian & Thick & Singles & $\sigma_{HG} = 0.066_{-0.023}^{+0.033}$ \\ 
\nodata & Thick & Multis & $\sigma_{HG} = 0.037_{-0.020}^{+0.038}$ \\
\nodata & Thin & Singles & $\sigma_{HG} = 0.025_{-0.007}^{+0.009}$ \\ 
\nodata & Thin & Multis & $\sigma_{HG} = 0.033_{-0.009}^{+0.013}$ \\ 
\hline
\enddata
\tablecomments{In all cases, the median of $\langle e \rangle$ is shown, and the uncertainties are calculated using the $16^{th}$ and $84^{th}$ distribution percentiles. The expected value of the half-Gaussian distribution is always zero by design. Therefore, we instead show the posterior median and uncertainties for $\sigma_{HG}$.}
\end{deluxetable*}

We find that, in the context of all four underlying eccentricity models, the eccentricities of singles in the kinematic thick disk are drawn from a distinct underlying distribution from singles in the kinematic thin disk. The data support a model in which single transiting planets in the kinematic thick disk have higher eccentricities, on average, than their kinematic thin-disk counterparts. However, we find no such evidence for multis: the underlying eccentricity distribution for multis in the kinematic thin disk is consistent with that of multis in the kinematic thick disk. 

\section{Discussion} \label{sec:discussion}

There are many stellar properties that change between the Kepler field stellar populations in the kinematic thin and thick disks. Thick disk stars are broadly older and more metal-poor than thin disk stars. Though we show in Figure \ref{fig:thick_thin_age_metal} that the difference between kinematic thick and thin disk metallicities within the Kepler sample is slight, we show that stellar age does vary significantly between the kinematic thin and thick disk samples. Since the stellar age—eccentricity relationship is not yet robustly understood, care is required to disentangle the effects of stellar age and kinematic thin/thick disk membership on orbital eccentricity. We leave the analysis of these more subtle relationships to future work. In the context of the current work, we demonstrate a marked difference between the eccentricities of singly-transiting planets in the kinematic thin versus thick disk, with thick disk singles being more eccentric. Such a trend reflects the opposite of the expectation from metallicity considerations alone. As the canonical picture goes, metal poorness (generally truer in the kinematic thick disk) ought to mean \textit{lower eccentricity} rather than higher \citep{dawson_giant_2013, buchhave_jupiter_2018}. This was first observed to be true among larger planets where the metallicity-eccentricity trends are starkest, but there exists evidence for the trend among smaller planets as well \citep{mills_california-kepler_2019}. In this context, we might doubly expect kinematic thick disk singles to have lower eccentricity: they are larger planets on average, of the size where the relative metal-poorness tends to manifest more clearly in lower eccentricities. Given that we observe the opposite trend, it is useful to carefully consider the potentially relevant physical mechanisms at work. Such a trend may well be driven by stellar age, close stellar encounters, other environmental effects, or a combination of these. 

\subsection{Singles vs. Multis}

As in previous works that investigate orbital eccentricities within a demographic framework \citep{van_eylen_orbital_2019, mills_california-kepler_2019, sagear_orbital_2023, gilbert_planets_2025, sagear_orbital_2025}, we elect to analyze singles and multis independently. It has been established that the orbital eccentricities of singles are drawn from an underlying distribution distinct from that of multis \citep{xie_exoplanet_2016, van_eylen_orbital_2019, mills_california-kepler_2019}. It is as yet uncertain whether this effect is diagnostic of two distinct pathways, or instead probing ends of the same continuous pebble-accretion story \citep{he_architectures_2020}. For example, observational singles could result from giant-impact dominated dynamical evolution, including planet-planet collisions in metastable original configurations \citep{pu__spacing_2015, chance_evidence_2024} while multis are dominated by photoevaporation throughout their lifetimes, and are not significantly dynamically disrupted, retaining their circular and compact configurations.

In the context of every model choice, we find that the difference in eccentricity between kinematic thin and thick disk planets is stronger in singles and hardly discernible for multis. There are several possible explanations for this effect. One potential reason for this difference is that multis must typically occupy more compact orbits to remain dynamically stable over time. Even a slight dynamical perturbation from e.g. a close stellar encounter \citep{schoettler_effect_2024, charalambous_breaking_2025} may disrupt the entire system architecture. Observational multis may reflect a survivorship bias of systems that have never been externally perturbed. Another possible explanation is related to the typical planet populations in multis vs. singles: planets within multi systems are much more likely to be small in radius than singles. Since there is a relationship between planet radius and eccentricity not only for small vs. large planets but also for planets in and outside of the radius gap \citep{gilbert_planets_2025, sagear_orbital_2025}, the planet populations themselves in the singles vs. multis sample may contribute to the strength of the kinematic thin vs. thick disk eccentricity difference. However, we present a preliminary check to assess the influence of giant planet occurrence on these results in Section \ref{subsec:giantplanets}. We find that our findings hold within the subsample of only small planets, so these results are unlikely to be caused to giant planet occurrence alone. What is needed to truly disentangle each of these effects is a comprehensive analysis along several axes (planet radius, stellar host metallicities and ages, Galactic kinematic properties and disk membership) and eccentricity.

\subsection{Control for Metallicity}

\begin{figure*}[ht!]
    \centering
    \includegraphics[width=\linewidth]{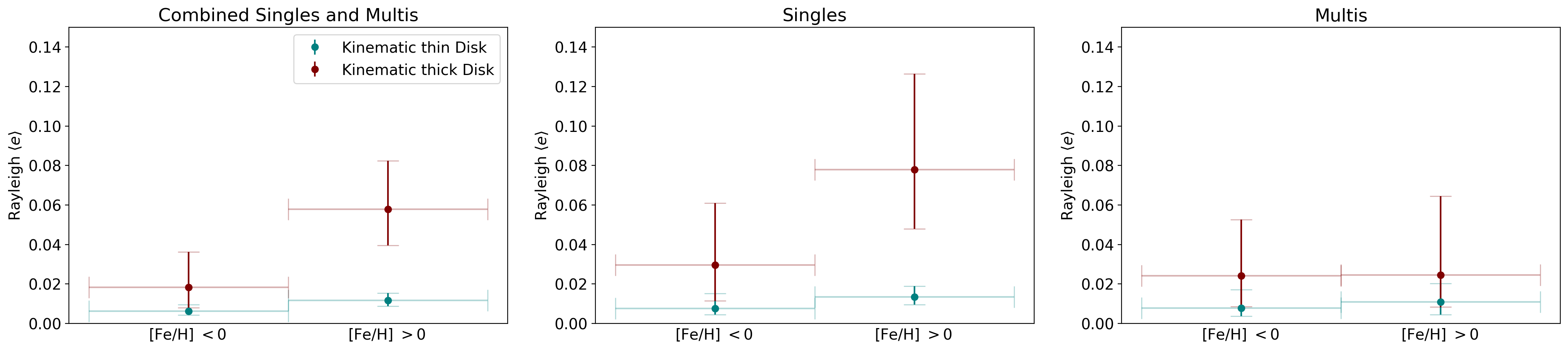}
    \caption{\textit{Left:} $\langle e \rangle$ for kinematic thick vs. thin disk planet hosts with $\mathrm{[Fe/H]} < 0$ and $\mathrm{[Fe/H]} > 0$. $\mathrm{[Fe/H]}$ values are taken from \citet{berger_gaia-kepler_2020}. $\langle e \rangle$ is calculated from the underlying eccentricity distribution for each subsample modeled as Rayleigh distributions. The vertical error bars represent the $1\sigma$ $\langle e \rangle$ uncertainties. \textit{Center:} Same as left panel, but for the single-transit sample. \textit{Right:} Same as left and center panels, but for the multi-transit sample.}
\label{fig:metalrichpoor}
\end{figure*}

We perform a cursory analysis of the role of kinematic disk membership vs. stellar metallicity on the orbital eccentricity. We split the sample up into two groups: stars with less than solar metallicity (``metal poor") and stars with more than solar metallicity (``metal rich"). We repeat the steps described in Section \ref{sec:methods} to fit the underlying eccentricity distribution for each group's kinematic thin and thick disk subpopulations, including the singles and the multis individually for each disk. Assuming a Rayleigh distribution, the underlying distributions are presented in Figure \ref{fig:metalrichpoor}. We find that for the singles, metal-rich kinematic thick disk planets have markedly higher eccentricities than metal-rich thin disk planets. However, this difference is not as apparent for the metal-poor sample. It has been previously shown that stellar metallicity may affect orbital eccentricity such that metal-rich stars tend to host more eccentric giant planets than their metal-poor analogues \citep[e.g.]{dawson_giant_2013}. Within this cursory analysis, the eccentricity—metallicity relation supports this trend; yet, the difference between kinematic thin and thick disk planet eccentricities is still apparent even for the metal-rich sample, suggesting that the eccentricity—disk membership trend is not fully explained by metallicity alone. Further work is needed to carefully disentangle the effects of metallicity, stellar age, and planet radius on orbital eccentricity to ultimately isolate the effect of each individual mechanism on the orbital dynamics of planets.

\subsection{Control for Giant Planet Occurrence}\label{subsec:giantplanets}
Elevated eccentricities in the kinematic thick disk appear to be dominated by the metal-rich ($[Fe/H]>0$) sample of singles (Figure \ref{fig:metalrichpoor}). It has been demonstrated that large planets with ($R_{p} > 3.5 R_{\oplus}$) exhibit systematically elevated eccentricities compared to smaller planets across several spectral types \citep{gilbert_planets_2025, sagear_orbital_2025}. It is also understood that the occurrence rate of eccentric planets is higher around metal-rich stellar hosts \citep{dawson_giant_2013}. It follows to question whether the elevated eccentricities in kinematic thick disk singles are simply covariant with giant planet occurrence.

\begin{figure*}[ht!]
    \centering
    \includegraphics[width=\linewidth]{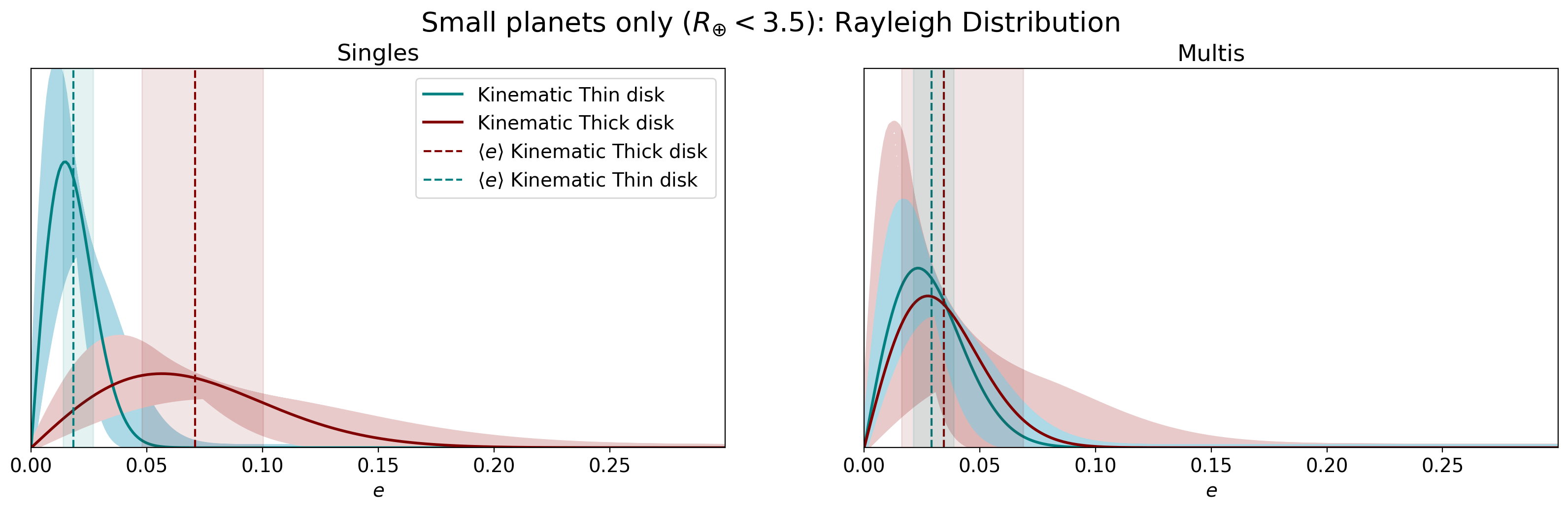}
    \caption{\textit{Left:} $\langle e \rangle$ for thick vs. thin disk small planet hosts with $R_p < 3.5 R_{\oplus}$. Singles are shown on the left panel, with multis on the right panel. $\langle e \rangle$ is calculated from the underlying eccentricity distribution for each subsample modeled as Rayleigh distributions. The vertical error bars represent the $1\sigma$ $\langle e \rangle$ uncertainties.}
\label{fig:smallplanets}
\end{figure*}

To test this, we repeat the analysis of kinematic thin vs. thick disk eccentricities with the Rayleigh distribution using only small planets with $R_{p} < 3.5 R_{\oplus}$. After filtering out all planets larger than $3.5 R_{\oplus}$, we have 437 thick disk planets (238 singles and 199 multis) and 1747 thin disk planets (961 singles and 786 multis.) We find elevated eccentricities to comparable significance in the kinematic thick disk sample of small planets (Figure \ref{fig:smallplanets}). This result suggests that to first order, the difference in $\langle e \rangle$ between the kinematic thin and thick disks is not fully explained by higher giant planet occurrence rate in the metal-rich singles sample alone.

\section{Conclusion} \label{sec:conclusion}

We summarize our work and results as follows:

\begin{itemize}
    
    \item We constrain orbital eccentricities for a sample of Kepler planet hosts across F, G, K, and M spectral types using Kepler transit light curves and stellar density priors from \citep{berger_gaia-kepler_2020}.
    \item We divide a calibration sample of Kepler—APOGEE stars into chemical thin disk and chemical thick disk subsamples. We kinematically determine thin and thick disk probabilities for the sample based on a GMM fit to the Galactocentric cylindrical velocities of each subsample.
    \item We extract the underlying eccentricity distribution within a Bayesian hierarchical framework, assuming alternately Rayleigh, Beta, monotonic Beta, and Half-Gaussian distributions. We perform a ``leave N out" validation to ensure our results are robust to individual eccentricity measurements. We find that thick disk single-transit systems tend to have higher eccentricities than thin disk single-transit planets.
    \item We perform a cursory analysis on the contribution of stellar host metallicity to the kinematic thick vs. thin disk eccentricity difference. We find that thick/thin disk eccentricities differ most strongly for planet hosts with $\mathrm{[Fe/H]} > 0$. We also perform a cursory analysis on the covariance of giant planet occurrence with elevated thick disk eccentricities, and we find that these results are not fully explained by giant planet occurrence alone.
    \item We orient these findings in the context of known eccentricity relations and consider the potential impact of stellar age, close stellar encounters, planetary atmospheric loss mechanisms, and various physical mechanisms that may link stellar and planetary orbital dynamics.
    
\end{itemize}

The presence of elevated eccentricities in the kinematic thick disk adds to a growing collection of evidence that planet occurrence and architectures appear to systematically reflect their Galactic environment, even above the first-order effects of host stellar conditions. Still, at this stage we cannot fully disentangle the relative effects of stellar host chemical properties (metallicity and $\alpha$-abundance), stellar age, Galactic dynamical properties, and planetary orbital eccentricities. Each of these physical properties and their potential effects on planet dynamics are intertwined with each another. In order to disentangle the relative effects of stellar properties and Galactic dynamics on planetary orbital eccentricities, a comprehensive analysis of each mechanism's individual contribution is necessary. A larger sample size of planet hosts in the kinematic thick disk would additionally strengthen those results.

Large scale transit surveys are forthcoming which plan for wider coverage of different dynamical areas of the Galaxy, such as the upcoming Roman Galactic Exoplanet Survey \citep{johnson_predictions_2020, wilson_transiting_2023} and Planetary Transits and Oscillations of stars (PLATO) mission \citep{rauer_plato_2025}. Combined with increased radial velocity coverage and astrometric planet occurrence information planned in Gaia DR4, a multi-axial analysis of stellar properties, Galactic dynamics, and planetary eccentricities is within reach and will reveal a detailed picture of late-stage planetary dynamical evolution.
    
\begin{acknowledgments}

We thank Natalia Guerrero, Quadry Chance, and Nazar Budaiev for discussions that improved the quality of this work.

This paper includes data collected by the \textit{Kepler} mission and obtained from the MAST data archive at the Space Telescope Science Institute (STScI). The specific observations analyzed can be accessed via \dataset[https://doi.org/10.17909/T9059R]{https://doi.org/10.17909/T9059R}. STScI is operated by the Association of Universities for Research in Astronomy, Inc., under NASA contract NAS5–26555. Support to MAST for these data is provided by the NASA Office of Space Science via grant NAG5–7584 and by other grants and contracts. This research has also made use of the NASA Exoplanet Archive \citep{nasa_exoplanet_archive_kepler_2019, christiansen_nasa_2025}, which is operated by the California Institute of Technology, under contract with the National Aeronautics and Space Administration under the Exoplanet Exploration Program.

This work has made use of data from the European Space Agency (ESA) mission {\it Gaia} (\url{https://www.cosmos.esa.int/gaia}), processed by the {\it Gaia} Data Processing and Analysis Consortium (DPAC, \url{https://www.cosmos.esa.int/web/gaia/dpac/consortium}). Funding for the DPAC has been provided by national institutions, in particular the institutions participating in the {\it Gaia} Multilateral Agreement. 

This publication makes use of data products from the Two Micron All Sky Survey, which is a joint project of the University of Massachusetts and the Infrared Processing and Analysis Center/California Institute of Technology, funded by the National Aeronautics and Space Administration and the National Science Foundation.

\end{acknowledgments}

\facilities{Kepler, Gaia, Sloan (APOGEE), Exoplanet Archive}

\software{numpy \citep{harris_array_2020}, matplotlib \citep{caswell_matplotlibmatplotlib_2024}, astropy \citep{robitaille_astropy_2013, collaboration_astropy_2018, collaboration_astropy_2022}, numpyro \citep{bingham_pyro_2019,phan_composable_2019}, dynesty \citep{skilling_nested_2004, skilling_nested_2006, higson_dynamic_2019, speagle_dynesty_2020, koposov_joshspeagledynesty_2024}, ALDERAAN \citep{gilbert_planets_2025}}

\bibliography{references}{}

\begin{thebibliography}{}
\expandafter\ifx\csname natexlab\endcsname\relax\def\natexlab#1{#1}\fi
\providecommand{\url}[1]{\href{#1}{#1}}
\providecommand{\dodoi}[1]{doi:~\href{http://doi.org/#1}{\nolinkurl{#1}}}
\providecommand{\doeprint}[1]{\href{http://ascl.net/#1}{\nolinkurl{http://ascl.net/#1}}}
\providecommand{\doarXiv}[1]{\href{https://arxiv.org/abs/#1}{\nolinkurl{https://arxiv.org/abs/#1}}}

\bibitem[{ {Abdurro’uf} {et~al.}(2022){Abdurro’uf}, Accetta, Aerts, Silva~Aguirre, Ahumada, Ajgaonkar, Filiz~Ak, Alam, Allende~Prieto, Almeida, Anders, Anderson, Andrews, Anguiano, Aquino-Ortíz, Aragón-Salamanca, Argudo-Fernández, Ata, Aubert, Avila-Reese, Badenes, Barbá, Barger, Barrera-Ballesteros, Beaton, Beers, Belfiore, Bender, Bernardi, Bershady, Beutler, Bidin, Bird, Bizyaev, Blanc, Blanton, Boardman, Bolton, Boquien, Borissova, Bovy, Brandt, Brown, Brownstein, Brusa, Buchner, Bundy, Burchett, Bureau, Burgasser, Cabang, Campbell, Cappellari, Carlberg, Wanderley, Carrera, Cash, Chen, Chen, Cherinka, Chiappini, Choi, Chojnowski, Chung, Clerc, Cohen, Comerford, Comparat, da~Costa, Covey, Crane, Cruz-Gonzalez, Culhane, Cunha, Dai, Damke, Darling, Davidson~Jr., Davies, Dawson, De~Lee, Diamond-Stanic, Cano-Díaz, Sánchez, Donor, Duckworth, Dwelly, Eisenstein, Elsworth, Emsellem, Eracleous, Escoffier, Fan, Farr, Feng, Fernández-Trincado, Feuillet, Filipp, Fillingham, Frinchaboy, Fromenteau, Galbany,
  García, García-Hernández, Ge, Geisler, Gelfand, Géron, Gibson, Goddy, Godoy-Rivera, Grabowski, Green, Greener, Grier, Griffith, Guo, Guy, Hadjara, Harding, Hasselquist, Hayes, Hearty, Hernández, Hill, Hogg, Holtzman, Horta, Hsieh, Hsu, Hsu, Huber, Huertas-Company, Hutchinson, Hwang, Ibarra-Medel, Chitham, Ilha, Imig, Jaekle, Jayasinghe, Ji, Johnson, Jones, Jönsson, Katkov, Khalatyan, Kinemuchi, Kisku, Knapen, Kneib, Kollmeier, Kong, Kounkel, Kreckel, Krishnarao, Lacerna, Lane, Langgin, Lavender, Law, Lazarz, Leung, Leung, Lewis, Li, Li, Lian, Liang, Lin, Lin, Lin, Lintott, Long, Longa-Peña, López-Cobá, Lu, Lundgren, Luo, Mackereth, de~la Macorra, Mahadevan, Majewski, Manchado, Mandeville, Maraston, Margalef-Bentabol, Masseron, Masters, Mathur, McDermid, Mckay, Merloni, Merrifield, Meszaros, Miglio, Di~Mille, Minniti, Minsley, Monachesi, Moon, Mosser, Mulchaey, Muna, Muñoz, Myers, Myers, Nadathur, Nair, Nandra, Neumann, Newman, Nidever, Nikakhtar, Nitschelm, O’Connell, Garma-Oehmichen, Luan
  Souza~de Oliveira, Olney, Oravetz, Ortigoza-Urdaneta, Osorio, Otter, Pace, Padilla, Pan, Pan, Parikh, Parker, Peirani, Peña~Ramírez, Penny, Percival, Perez-Fournon, Pinsonneault, Poidevin, Poovelil, Price-Whelan, Bárbara~de Andrade~Queiroz, Raddick, Ray, Rembold, Riddle, Riffel, Riffel, Rix, Robin, Rodríguez-Puebla, Roman-Lopes, Román-Zúñiga, Rose, Ross, Rossi, Rubin, Salvato, Sánchez, Sánchez-Gallego, Sanderson, Santana~Rojas, Sarceno, Sarmiento, Sayres, Sazonova, Schaefer, Schiavon, Schlegel, Schneider, Schultheis, Schwope, Serenelli, Serna, Shao, Shapiro, Sharma, Shen, Shetrone, Shu, Simon, Skrutskie, Smethurst, Smith, Sobeck, Spoo, Sprague, Stark, Stassun, Steinmetz, Stello, Stone-Martinez, Storchi-Bergmann, Stringfellow, Stutz, Su, Taghizadeh-Popp, Talbot, Tayar, Telles, Teske, Thakar, Theissen, Tkachenko, Thomas, Tojeiro, Hernandez~Toledo, Troup, Trump, Trussler, Turner, Tuttle, Unda-Sanzana, Vázquez-Mata, Valentini, Valenzuela, Vargas-González, Vargas-Magaña, Alfaro, Villanova, Vincenzo,
  Wake, Warfield, Washington, Weaver, Weijmans, Weinberg, Weiss, Westfall, Wild, Wilde, Wilson, Wilson, Wilson, Wolf, Wood-Vasey, Yan, Zamora, Zasowski, Zhang, Zhao, Zheng, Zheng, \& Zhu}]{abdurrouf_seventeenth_2022}
{Abdurro’uf}, Accetta, K., Aerts, C., {et~al.} 2022, \bibinfo{title}{The {Seventeenth} {Data} {Release} of the {Sloan} {Digital} {Sky} {Surveys}: {Complete} {Release} of {MaNGA}, {MaStar}, and {APOGEE}-2 {Data},} The Astrophysical Journal Supplement Series, 259, 35, \dodoi{10.3847/1538-4365/ac4414}

\bibitem[{S.~G. Anderson {et~al.}(2021)Anderson, Dittmann, Ballard, \& Bedell}]{anderson_higher_2021}
Anderson, S.~G., Dittmann, J.~A., Ballard, S., \& Bedell, M. 2021, \bibinfo{title}{Higher {Compact} {Multiple} {Occurrence} around {Metal}-poor {M}-dwarfs and {Late}-{K}-dwarfs,} The Astronomical Journal, 161, 203, \dodoi{10.3847/1538-3881/abe70b}

\bibitem[{R. Angus {et~al.}(2022)Angus, Price-Whelan, Zinn, Bedell, Lu, \& Foreman-Mackey}]{angus_3d_2022}
Angus, R., Price-Whelan, A.~M., Zinn, J.~C., {et~al.} 2022, \bibinfo{title}{The {3D} {Galactocentric} {Velocities} of {Kepler} {Stars}: {Marginalizing} {Over} {Missing} {Radial} {Velocities},} The Astronomical Journal, 164, 25, \dodoi{10.3847/1538-3881/ac6fea}

\bibitem[{S.~A. Barnes(2007)Barnes}]{barnes_ages_2007}
Barnes, S.~A. 2007, \bibinfo{title}{Ages for illustrative field stars using gyrochronology: viability, limitations and errors,} The Astrophysical Journal, 669, 1167, \dodoi{10.1086/519295}

\bibitem[{D. Bashi \& S. Zucker(2022)Bashi \& Zucker}]{bashi_exoplanets_2022}
Bashi, D., \& Zucker, S. 2022, \bibinfo{title}{Exoplanets in the {Galactic} context: planet occurrence rates in the thin disc, thick disc, and stellar halo of {Kepler} stars,} Monthly Notices of the Royal Astronomical Society, 510, 3449, \dodoi{10.1093/mnras/stab3596}

\bibitem[{R.~L. Beaton {et~al.}(2021)Beaton, Oelkers, Hayes, Covey, Chojnowski, De~Lee, Sobeck, Majewski, Cohen, Fernández-Trincado, Longa-Peña, O’Connell, Santana, Stringfellow, Zasowski, Aerts, Anguiano, Bender, Cañas, Cunha, Donor, Fleming, Frinchaboy, Feuillet, Harding, Hasselquist, Holtzman, Johnson, Kollmeier, Kounkel, Mahadevan, Price-Whelan, Rojas-Arriagada, Román-Zúñiga, Schlafly, Schultheis, Shetrone, Simon, Stassun, Stutz, Tayar, Teske, Tkachenko, Troup, Albareti, Bizyaev, Bovy, Burgasser, Comparat, Downes, Geisler, Inno, Manchado, Ness, Pinsonneault, Prada, Roman-Lopes, Simonian, Smith, Yan, \& Zamora}]{beaton_final_2021}
Beaton, R.~L., Oelkers, R.~J., Hayes, C.~R., {et~al.} 2021, \bibinfo{title}{Final {Targeting} {Strategy} for the {Sloan} {Digital} {Sky} {Survey} {IV} {Apache} {Point} {Observatory} {Galactic} {Evolution} {Experiment} 2 {North} {Survey},} The Astronomical Journal, 162, 302, \dodoi{10.3847/1538-3881/ac260c}

\bibitem[{T. Bensby {et~al.}(2003)Bensby, Feltzing, \& Lundström}]{bensby_elemental_2003}
Bensby, T., Feltzing, S., \& Lundström, I. 2003, \bibinfo{title}{Elemental abundance trends in the {Galactic} thin and thick disks as traced by nearby {F} and {G} dwarf stars,} Astronomy \& Astrophysics, 410, 527, \dodoi{10.1051/0004-6361:20031213}

\bibitem[{T. Bensby {et~al.}(2014)Bensby, Feltzing, \& Oey}]{bensby_exploring_2014}
Bensby, T., Feltzing, S., \& Oey, M.~S. 2014, \bibinfo{title}{Exploring the {Milky} {Way} stellar disk - {A} detailed elemental abundance study of 714 {F} and {G} dwarf stars in the solar neighbourhood,} Astronomy \& Astrophysics, 562, A71, \dodoi{10.1051/0004-6361/201322631}

\bibitem[{T.~A. Berger {et~al.}(2018)Berger, Huber, Gaidos, \& van Saders}]{berger_revised_2018}
Berger, T.~A., Huber, D., Gaidos, E., \& van Saders, J.~L. 2018, \bibinfo{title}{Revised {Radii} of \textit{{Kepler}} {Stars} and {Planets} {Using} \textit{{Gaia}} {Data} {Release} 2,} The Astrophysical Journal, 866, 99, \dodoi{10.3847/1538-4357/aada83}

\bibitem[{T.~A. Berger {et~al.}(2020)Berger, Huber, Gaidos, van Saders, \& Weiss}]{berger_gaia-kepler_2020}
Berger, T.~A., Huber, D., Gaidos, E., van Saders, J.~L., \& Weiss, L.~M. 2020, \bibinfo{title}{The {Gaia}-{Kepler} {Stellar} {Properties} {Catalog}. {II}. {Planet} {Radius} {Demographics} as a {Function} of {Stellar} {Mass} and {Age},} The Astronomical Journal, 160, 108, \dodoi{10.3847/1538-3881/aba18a}

\bibitem[{M.~J. Betancourt \& M. Girolami(2013)Betancourt \& Girolami}]{betancourt_hamiltonian_2013}
Betancourt, M.~J., \& Girolami, M. 2013, Hamiltonian {Monte} {Carlo} for {Hierarchical} {Models}, arXiv, \dodoi{10.48550/arXiv.1312.0906}

\bibitem[{E. Bingham {et~al.}(2019)Bingham, Chen, Jankowiak, Obermeyer, Pradhan, Karaletsos, Singh, Szerlip, Horsfall, \& Goodman}]{bingham_pyro_2019}
Bingham, E., Chen, J.~P., Jankowiak, M., {et~al.} 2019, \bibinfo{title}{Pyro: {Deep} {Universal} {Probabilistic} {Programming},} Journal of Machine Learning Research, 20, 1.
\newblock \url{http://jmlr.org/papers/v20/18-403.html}

\bibitem[{W.~J. Borucki {et~al.}(2010)Borucki, Koch, Basri, Batalha, Brown, Caldwell, Caldwell, Christensen-Dalsgaard, Cochran, DeVore, Dunham, Dupree, Gautier, Geary, Gilliland, Gould, Howell, Jenkins, Kondo, Latham, Marcy, Meibom, Kjeldsen, Lissauer, Monet, Morrison, Sasselov, Tarter, Boss, Brownlee, Owen, Buzasi, Charbonneau, Doyle, Fortney, Ford, Holman, Seager, Steffen, Welsh, Rowe, Anderson, Buchhave, Ciardi, Walkowicz, Sherry, Horch, Isaacson, Everett, Fischer, Torres, Johnson, Endl, MacQueen, Bryson, Dotson, Haas, Kolodziejczak, Van~Cleve, Chandrasekaran, Twicken, Quintana, Clarke, Allen, Li, Wu, Tenenbaum, Verner, Bruhweiler, Barnes, \& Prsa}]{borucki_kepler_2010}
Borucki, W.~J., Koch, D., Basri, G., {et~al.} 2010, \bibinfo{title}{Kepler {Planet}-{Detection} {Mission}: {Introduction} and {First} {Results},} Science, 327, 977, \dodoi{10.1126/science.1185402}

\bibitem[{J.~M. Brewer {et~al.}(2018)Brewer, Wang, Fischer, \& Foreman-Mackey}]{brewer_compact_2018}
Brewer, J.~M., Wang, S., Fischer, D.~A., \& Foreman-Mackey, D. 2018, \bibinfo{title}{Compact {Multi}-planet {Systems} are more {Common} around {Metal}-poor {Hosts},} The Astrophysical Journal Letters, 867, L3, \dodoi{10.3847/2041-8213/aae710}

\bibitem[{L.~A. Buchhave {et~al.}(2018)Buchhave, Bitsch, Johansen, Latham, Bizzarro, Bieryla, \& Kipping}]{buchhave_jupiter_2018}
Buchhave, L.~A., Bitsch, B., Johansen, A., {et~al.} 2018, \bibinfo{title}{Jupiter {Analogs} {Orbit} {Stars} with an {Average} {Metallicity} {Close} to {That} of the {Sun},} The Astrophysical Journal, 856, 37, \dodoi{10.3847/1538-4357/aaafca}

\bibitem[{C.~J. Burke(2008)Burke}]{burke_impact_2008}
Burke, C.~J. 2008, \bibinfo{title}{Impact of {Orbital} {Eccentricity} on the {Detection} of {Transiting} {Extrasolar} {Planets},} The Astrophysical Journal, 679, 1566, \dodoi{10.1086/587798}

\bibitem[{T.~A. Caswell {et~al.}(2024)Caswell, Andrade, Lee, Droettboom, Hoffmann, Klymak, Hunter, Firing, Stansby, Varoquaux, Nielsen, Gustafsson, Sunden, Root, May, hannah, Elson, Seppänen, Lee, Dale, McDougall, Straw, Lucas, Hobson, Comer, Gohlke, Vincent, Yu, Ma, \& Silvester}]{caswell_matplotlibmatplotlib_2024}
Caswell, T.~A., Andrade, E. S.~d., Lee, A., {et~al.} 2024, matplotlib/matplotlib: {REL}: v3.7.5, Zenodo, \dodoi{10.5281/zenodo.10669804}

\bibitem[{Q. Chance \& S. Ballard(2024)Chance \& Ballard}]{chance_evidence_2024}
Chance, Q., \& Ballard, S. 2024, Evidence that {Planets} in the {Radius} {Gap} {Do} {Not} {Resemble} {Their} {Neighbors}, arXiv, \dodoi{10.48550/arXiv.2410.02150}

\bibitem[{C. Charalambous {et~al.}(2025)Charalambous, Cuello, \& Petrovich}]{charalambous_breaking_2025}
Charalambous, C., Cuello, N., \& Petrovich, C. 2025, \bibinfo{title}{Breaking long-period resonance chains with stellar flybys,} Astronomy \& Astrophysics, 696, A175, \dodoi{10.1051/0004-6361/202553710}

\bibitem[{D.-C. Chen {et~al.}(2021)Chen, Xie, Zhou, Dong, Liu, Wang, Xiang, Huang, Luo, \& Zheng}]{chen_planets_2021}
Chen, D.-C., Xie, J.-W., Zhou, J.-L., {et~al.} 2021, \bibinfo{title}{Planets {Across} {Space} and {Time} ({PAST}). {I}. {Characterizing} the {Memberships} of {Galactic} {Components} and {Stellar} {Ages}: {Revisiting} the {Kinematic} {Methods} and {Applying} to {Planet} {Host} {Stars},} The Astrophysical Journal, 909, 115, \dodoi{10.3847/1538-4357/abd5be}

\bibitem[{J.~L. Christiansen {et~al.}(2025)Christiansen, McElroy, Harbut, Ciardi, Crane, Good, Hardegree-Ullman, Kesseli, Lund, Lynn, Muthiar, Nilsson, Oluyide, Papin, Rivera, Swain, Susemiehl, Tam, Eyken, \& Beichman}]{christiansen_nasa_2025}
Christiansen, J.~L., McElroy, D.~L., Harbut, M., {et~al.} 2025, \bibinfo{title}{The {NASA} {Exoplanet} {Archive} and {Exoplanet} {Follow}-up {Observing} {Program}: {Data}, {Tools}, and {Usage},} The Planetary Science Journal, 6, 186, \dodoi{10.3847/PSJ/ade3c2}

\bibitem[{T.~A. Collaboration {et~al.}(2018)Collaboration, Price-Whelan, Sipőcz, Günther, Lim, Crawford, Conseil, Shupe, Craig, Dencheva, Ginsburg, VanderPlas, Bradley, Pérez-Suárez, Val-Borro, Contributors), Aldcroft, Cruz, Robitaille, Tollerud, Committee), Ardelean, Babej, Bach, Bachetti, Bakanov, Bamford, Barentsen, Barmby, Baumbach, Berry, Biscani, Boquien, Bostroem, Bouma, Brammer, Bray, Breytenbach, Buddelmeijer, Burke, Calderone, Rodríguez, Cara, Cardoso, Cheedella, Copin, Corrales, Crichton, D’Avella, Deil, Depagne, Dietrich, Donath, Droettboom, Earl, Erben, Fabbro, Ferreira, Finethy, Fox, Garrison, Gibbons, Goldstein, Gommers, Greco, Greenfield, Groener, Grollier, Hagen, Hirst, Homeier, Horton, Hosseinzadeh, Hu, Hunkeler, Ivezić, Jain, Jenness, Kanarek, Kendrew, Kern, Kerzendorf, Khvalko, King, Kirkby, Kulkarni, Kumar, Lee, Lenz, Littlefair, Ma, Macleod, Mastropietro, McCully, Montagnac, Morris, Mueller, Mumford, Muna, Murphy, Nelson, Nguyen, Ninan, Nöthe, Ogaz, Oh, Parejko, Parley, Pascual,
  Patil, Patil, Plunkett, Prochaska, Rastogi, Janga, Sabater, Sakurikar, Seifert, Sherbert, Sherwood-Taylor, Shih, Sick, Silbiger, Singanamalla, Singer, Sladen, Sooley, Sornarajah, Streicher, Teuben, Thomas, Tremblay, Turner, Terrón, Kerkwijk, Vega, Watkins, Weaver, Whitmore, Woillez, Zabalza, \& Contributors)}]{collaboration_astropy_2018}
Collaboration, T.~A., Price-Whelan, A.~M., Sipőcz, B.~M., {et~al.} 2018, \bibinfo{title}{The {Astropy} {Project}: {Building} an {Open}-science {Project} and {Status} of the v2.0 {Core} {Package}*,} The Astronomical Journal, 156, 123, \dodoi{10.3847/1538-3881/aabc4f}

\bibitem[{T.~A. Collaboration {et~al.}(2022)Collaboration, Price-Whelan, Lim, Earl, Starkman, Bradley, Shupe, Patil, Corrales, Brasseur, Nöthe, Donath, Tollerud, Morris, Ginsburg, Vaher, Weaver, Tocknell, Jamieson, Kerkwijk, Robitaille, Merry, Bachetti, Günther, Authors, Aldcroft, Alvarado-Montes, Archibald, Bódi, Bapat, Barentsen, Bazán, Biswas, Boquien, Burke, Cara, Cara, Conroy, Conseil, Craig, Cross, Cruz, D’Eugenio, Dencheva, Devillepoix, Dietrich, Eigenbrot, Erben, Ferreira, Foreman-Mackey, Fox, Freij, Garg, Geda, Glattly, Gondhalekar, Gordon, Grant, Greenfield, Groener, Guest, Gurovich, Handberg, Hart, Hatfield-Dodds, Homeier, Hosseinzadeh, Jenness, Jones, Joseph, Kalmbach, Karamehmetoglu, Kałuszyński, Kelley, Kern, Kerzendorf, Koch, Kulumani, Lee, Ly, Ma, MacBride, Maljaars, Muna, Murphy, Norman, O’Steen, Oman, Pacifici, Pascual, Pascual-Granado, Patil, Perren, Pickering, Rastogi, Roulston, Ryan, Rykoff, Sabater, Sakurikar, Salgado, Sanghi, Saunders, Savchenko, Schwardt, Seifert-Eckert,
  Shih, Jain, Shukla, Sick, Simpson, Singanamalla, Singer, Singhal, Sinha, Sipőcz, Spitler, Stansby, Streicher, Šumak, Swinbank, Taranu, Tewary, Tremblay, Val-Borro, Kooten, Vasović, Verma, Cardoso, Williams, Wilson, Winkel, Wood-Vasey, Xue, Yoachim, Zhang, Zonca, \& Contributors}]{collaboration_astropy_2022}
Collaboration, T.~A., Price-Whelan, A.~M., Lim, P.~L., {et~al.} 2022, \bibinfo{title}{The {Astropy} {Project}: {Sustaining} and {Growing} a {Community}-oriented {Open}-source {Project} and the {Latest} {Major} {Release} (v5.0) of the {Core} {Package}*,} The Astrophysical Journal, 935, 167, \dodoi{10.3847/1538-4357/ac7c74}

\bibitem[{J.~A. Correa-Otto \& R.~A. Gil-Hutton(2017)Correa-Otto \& Gil-Hutton}]{correa-otto_galactic_2017}
Correa-Otto, J.~A., \& Gil-Hutton, R.~A. 2017, \bibinfo{title}{Galactic perturbations on the population of wide binary stars with exoplanets,} Astronomy \& Astrophysics, 608, A116, \dodoi{10.1051/0004-6361/201731229}

\bibitem[{R.~I. Dawson \& J.~A. Johnson(2012)Dawson \& Johnson}]{dawson_photoeccentric_2012}
Dawson, R.~I., \& Johnson, J.~A. 2012, \bibinfo{title}{{THE} {PHOTOECCENTRIC} {EFFECT} {AND} {PROTO}-{HOT} {JUPITERS}. {I}. {MEASURING} {PHOTOMETRIC} {ECCENTRICITIES} {OF} {INDIVIDUAL} {TRANSITING} {PLANETS},} The Astrophysical Journal, 756, 122, \dodoi{10.1088/0004-637X/756/2/122}

\bibitem[{R.~I. Dawson {et~al.}(2016)Dawson, Lee, \& Chiang}]{dawson_correlations_2016}
Dawson, R.~I., Lee, E.~J., \& Chiang, E. 2016, \bibinfo{title}{{CORRELATIONS} {BETWEEN} {COMPOSITIONS} {AND} {ORBITS} {ESTABLISHED} {BY} {THE} {GIANT} {IMPACT} {ERA} {OF} {PLANET} {FORMATION},} The Astrophysical Journal, 822, 54, \dodoi{10.3847/0004-637X/822/1/54}

\bibitem[{R.~I. Dawson \& R.~A. Murray-Clay(2013)Dawson \& Murray-Clay}]{dawson_giant_2013}
Dawson, R.~I., \& Murray-Clay, R.~A. 2013, \bibinfo{title}{{GIANT} {PLANETS} {ORBITING} {METAL}-{RICH} {STARS} {SHOW} {SIGNATURES} {OF} {PLANET}-{PLANET} {INTERACTIONS},} The Astrophysical Journal, 767, L24, \dodoi{10.1088/2041-8205/767/2/L24}

\bibitem[{C.~R. Do~Ó {et~al.}(2023)Do~Ó, O’Neil, Konopacky, Do, Martinez, Ruffio, \& Ghez}]{do_o_orbital_2023}
Do~Ó, C.~R., O’Neil, K.~K., Konopacky, Q.~M., {et~al.} 2023, \bibinfo{title}{The {Orbital} {Eccentricities} of {Directly} {Imaged} {Companions} {Using} {Observable}-based {Priors}: {Implications} for {Population}-level {Distributions},} The Astronomical Journal, 166, 48, \dodoi{10.3847/1538-3881/acdc9a}

\bibitem[{E.~B. Ford {et~al.}(2008)Ford, Quinn, \& Veras}]{ford_characterizing_2008}
Ford, E.~B., Quinn, S.~N., \& Veras, D. 2008, \bibinfo{title}{Characterizing the {Orbital} {Eccentricities} of {Transiting} {Extrasolar} {Planets} with {Photometric} {Observations},} The Astrophysical Journal, 678, 1407, \dodoi{10.1086/587046}

\bibitem[{D. Foreman-Mackey {et~al.}(2021)Foreman-Mackey, Luger, Agol, Barclay, Bouma, Brandt, Czekala, David, Dong, Gilbert, Gordon, Hedges, Hey, Morris, Price-Whelan, \& Savel}]{foreman-mackey_exoplanet_2021}
Foreman-Mackey, D., Luger, R., Agol, E., {et~al.} 2021, \bibinfo{title}{`exoplanet`: {Gradient}-based probabilistic inference for exoplanet data \& other astronomical time series,} Journal of Open Source Software, 6, 3285, \dodoi{10.21105/joss.03285}

\bibitem[{B.~J. Fulton \& E.~A. Petigura(2018)Fulton \& Petigura}]{fulton_california-kepler_2018}
Fulton, B.~J., \& Petigura, E.~A. 2018, \bibinfo{title}{The {California}-{Kepler} {Survey}. {VII}. {Precise} {Planet} {Radii} {Leveraging} {Gaia} {DR2} {Reveal} the {Stellar} {Mass} {Dependence} of the {Planet} {Radius} {Gap},} The Astronomical Journal, 156, 264, \dodoi{10.3847/1538-3881/aae828}

\bibitem[{E. Furlan {et~al.}(2017)Furlan, Ciardi, Everett, Saylors, Teske, Horch, Howell, Belle, Hirsch, Gautier, Adams, Barrado, Cartier, Dressing, Dupree, Gilliland, Lillo-Box, Lucas, \& Wang}]{furlan_lessigreaterkeplerlessigreaterfollow-up_2017}
Furlan, E., Ciardi, D.~R., Everett, M.~E., {et~al.} 2017, \bibinfo{title}{{THE}\${\textbackslash}less\$i\${\textbackslash}greater\${KEPLER}\${\textbackslash}less\$/i\${\textbackslash}greater\${FOLLOW}-{UP} {OBSERVATION} {PROGRAM}. {I}. {A} {CATALOG} {OF} {COMPANIONS} {TO}\${\textbackslash}less\$i\${\textbackslash}greater\${KEPLER}\${\textbackslash}less\$/i\${\textbackslash}greater\${STARS} {FROM} {HIGH}-{RESOLUTION} {IMAGING},} The Astronomical Journal, 153, 71, \dodoi{10.3847/1538-3881/153/2/71}

\bibitem[{ {Gaia Collaboration} {et~al.}(2016){Gaia Collaboration}, {Prusti, T.}, {de Bruijne, J. H. J.}, {Brown, A. G. A.}, {Vallenari, A.}, {Babusiaux, C.}, {Bailer-Jones, C. A. L.}, {Bastian, U.}, {Biermann, M.}, {Evans, D. W.}, {Eyer, L.}, {Jansen, F.}, {Jordi, C.}, {Klioner, S. A.}, {Lammers, U.}, {Lindegren, L.}, {Luri, X.}, {Mignard, F.}, {Milligan, D. J.}, {Panem, C.}, {Poinsignon, V.}, {Pourbaix, D.}, {Randich, S.}, {Sarri, G.}, {Sartoretti, P.}, {Siddiqui, H. I.}, {Soubiran, C.}, {Valette, V.}, {van Leeuwen, F.}, {Walton, N. A.}, {Aerts, C.}, {Arenou, F.}, {Cropper, M.}, {Drimmel, R.}, {Høg, E.}, {Katz, D.}, {Lattanzi, M. G.}, {O{\textbackslash}'Mullane, W.}, {Grebel, E. K.}, {Holland, A. D.}, {Huc, C.}, {Passot, X.}, {Bramante, L.}, {Cacciari, C.}, {Castañeda, J.}, {Chaoul, L.}, {Cheek, N.}, {De Angeli, F.}, {Fabricius, C.}, {Guerra, R.}, {Hernández, J.}, {Jean-Antoine-Piccolo, A.}, {Masana, E.}, {Messineo, R.}, {Mowlavi, N.}, {Nienartowicz, K.}, {Ordóñez-Blanco, D.}, {Panuzzo, P.}, {Portell,
  J.}, {Richards, P. J.}, {Riello, M.}, {Seabroke, G. M.}, {Tanga, P.}, {Thévenin, F.}, {Torra, J.}, {Els, S. G.}, {Gracia-Abril, G.}, {Comoretto, G.}, {Garcia-Reinaldos, M.}, {Lock, T.}, {Mercier, E.}, {Altmann, M.}, {Andrae, R.}, {Astraatmadja, T. L.}, {Bellas-Velidis, I.}, {Benson, K.}, {Berthier, J.}, {Blomme, R.}, {Busso, G.}, {Carry, B.}, {Cellino, A.}, {Clementini, G.}, {Cowell, S.}, {Creevey, O.}, {Cuypers, J.}, {Davidson, M.}, {De Ridder, J.}, {de Torres, A.}, {Delchambre, L.}, {Dell{\textbackslash}'Oro, A.}, {Ducourant, C.}, {Frémat, Y.}, {García-Torres, M.}, {Gosset, E.}, {Halbwachs, J.-L.}, {Hambly, N. C.}, {Harrison, D. L.}, {Hauser, M.}, {Hestroffer, D.}, {Hodgkin, S. T.}, {Huckle, H. E.}, {Hutton, A.}, {Jasniewicz, G.}, {Jordan, S.}, {Kontizas, M.}, {Korn, A. J.}, {Lanzafame, A. C.}, {Manteiga, M.}, {Moitinho, A.}, {Muinonen, K.}, {Osinde, J.}, {Pancino, E.}, {Pauwels, T.}, {Petit, J.-M.}, {Recio-Blanco, A.}, {Robin, A. C.}, {Sarro, L. M.}, {Siopis, C.}, {Smith, M.}, {Smith, K. W.},
  {Sozzetti, A.}, {Thuillot, W.}, {van Reeven, W.}, {Viala, Y.}, {Abbas, U.}, {Abreu Aramburu, A.}, {Accart, S.}, {Aguado, J. J.}, {Allan, P. M.}, {Allasia, W.}, {Altavilla, G.}, {Álvarez, M. A.}, {Alves, J.}, {Anderson, R. I.}, {Andrei, A. H.}, {Anglada Varela, E.}, {Antiche, E.}, {Antoja, T.}, {Antón, S.}, {Arcay, B.}, {Atzei, A.}, {Ayache, L.}, {Bach, N.}, {Baker, S. G.}, {Balaguer-Núñez, L.}, {Barache, C.}, {Barata, C.}, {Barbier, A.}, {Barblan, F.}, {Baroni, M.}, {Barrado y Navascués, D.}, {Barros, M.}, {Barstow, M. A.}, {Becciani, U.}, {Bellazzini, M.}, {Bellei, G.}, {Bello García, A.}, {Belokurov, V.}, {Bendjoya, P.}, {Berihuete, A.}, {Bianchi, L.}, {Bienaymé, O.}, {Billebaud, F.}, {Blagorodnova, N.}, {Blanco-Cuaresma, S.}, {Boch, T.}, {Bombrun, A.}, {Borrachero, R.}, {Bouquillon, S.}, {Bourda, G.}, {Bouy, H.}, {Bragaglia, A.}, {Breddels, M. A.}, {Brouillet, N.}, {Brüsemeister, T.}, {Bucciarelli, B.}, {Budnik, F.}, {Burgess, P.}, {Burgon, R.}, {Burlacu, A.}, {Busonero, D.}, {Buzzi, R.},
  {Caffau, E.}, {Cambras, J.}, {Campbell, H.}, {Cancelliere, R.}, {Cantat-Gaudin, T.}, {Carlucci, T.}, {Carrasco, J. M.}, {Castellani, M.}, {Charlot, P.}, {Charnas, J.}, {Charvet, P.}, {Chassat, F.}, {Chiavassa, A.}, {Clotet, M.}, {Cocozza, G.}, {Collins, R. S.}, {Collins, P.}, {Costigan, G.}, {Crifo, F.}, {Cross, N. J. G.}, {Crosta, M.}, {Crowley, C.}, {Dafonte, C.}, {Damerdji, Y.}, {Dapergolas, A.}, {David, P.}, {David, M.}, {De Cat, P.}, {de Felice, F.}, {de Laverny, P.}, {De Luise, F.}, {De March, R.}, {de Martino, D.}, {de Souza, R.}, {Debosscher, J.}, {del Pozo, E.}, {Delbo, M.}, {Delgado, A.}, {Delgado, H. E.}, {di Marco, F.}, {Di Matteo, P.}, {Diakite, S.}, {Distefano, E.}, {Dolding, C.}, {Dos Anjos, S.}, {Drazinos, P.}, {Durán, J.}, {Dzigan, Y.}, {Ecale, E.}, {Edvardsson, B.}, {Enke, H.}, {Erdmann, M.}, {Escolar, D.}, {Espina, M.}, {Evans, N. W.}, {Eynard Bontemps, G.}, {Fabre, C.}, {Fabrizio, M.}, {Faigler, S.}, {Falcão, A. J.}, {Farràs Casas, M.}, {Faye, F.}, {Federici, L.}, {Fedorets, G.},
  {Fernández-Hernández, J.}, {Fernique, P.}, {Fienga, A.}, {Figueras, F.}, {Filippi, F.}, {Findeisen, K.}, {Fonti, A.}, {Fouesneau, M.}, {Fraile, E.}, {Fraser, M.}, {Fuchs, J.}, {Furnell, R.}, {Gai, M.}, {Galleti, S.}, {Galluccio, L.}, {Garabato, D.}, {García-Sedano, F.}, {Garé, P.}, {Garofalo, A.}, {Garralda, N.}, {Gavras, P.}, {Gerssen, J.}, {Geyer, R.}, {Gilmore, G.}, {Girona, S.}, {Giuffrida, G.}, {Gomes, M.}, {González-Marcos, A.}, {González-Núñez, J.}, {González-Vidal, J. J.}, {Granvik, M.}, {Guerrier, A.}, {Guillout, P.}, {Guiraud, J.}, {Gúrpide, A.}, {Gutiérrez-Sánchez, R.}, {Guy, L. P.}, {Haigron, R.}, {Hatzidimitriou, D.}, {Haywood, M.}, {Heiter, U.}, {Helmi, A.}, {Hobbs, D.}, {Hofmann, W.}, {Holl, B.}, {Holland, G.}, {Hunt, J. A. S.}, {Hypki, A.}, {Icardi, V.}, {Irwin, M.}, {Jevardat de Fombelle, G.}, {Jofré, P.}, {Jonker, P. G.}, {Jorissen, A.}, {Julbe, F.}, {Karampelas, A.}, {Kochoska, A.}, {Kohley, R.}, {Kolenberg, K.}, {Kontizas, E.}, {Koposov, S. E.}, {Kordopatis, G.}, {Koubsky,
  P.}, {Kowalczyk, A.}, {Krone-Martins, A.}, {Kudryashova, M.}, {Kull, I.}, {Bachchan, R. K.}, {Lacoste-Seris, F.}, {Lanza, A. F.}, {Lavigne, J.-B.}, {Le Poncin-Lafitte, C.}, {Lebreton, Y.}, {Lebzelter, T.}, {Leccia, S.}, {Leclerc, N.}, {Lecoeur-Taibi, I.}, {Lemaitre, V.}, {Lenhardt, H.}, {Leroux, F.}, {Liao, S.}, {Licata, E.}, {Lindstrøm, H. E. P.}, {Lister, T. A.}, {Livanou, E.}, {Lobel, A.}, {Löffler, W.}, {López, M.}, {Lopez-Lozano, A.}, {Lorenz, D.}, {Loureiro, T.}, {MacDonald, I.}, {Magalhães Fernandes, T.}, {Managau, S.}, {Mann, R. G.}, {Mantelet, G.}, {Marchal, O.}, {Marchant, J. M.}, {Marconi, M.}, {Marie, J.}, {Marinoni, S.}, {Marrese, P. M.}, {Marschalkó, G.}, {Marshall, D. J.}, {Martín-Fleitas, J. M.}, {Martino, M.}, {Mary, N.}, {Matijevic, G.}, {Mazeh, T.}, {McMillan, P. J.}, {Messina, S.}, {Mestre, A.}, {Michalik, D.}, {Millar, N. R.}, {Miranda, B. M. H.}, {Molina, D.}, {Molinaro, R.}, {Molinaro, M.}, {Molnár, L.}, {Moniez, M.}, {Montegriffo, P.}, {Monteiro, D.}, {Mor, R.}, {Mora, A.},
  {Morbidelli, R.}, {Morel, T.}, {Morgenthaler, S.}, {Morley, T.}, {Morris, D.}, {Mulone, A. F.}, {Muraveva, T.}, {Musella, I.}, {Narbonne, J.}, {Nelemans, G.}, {Nicastro, L.}, {Noval, L.}, {Ordénovic, C.}, {Ordieres-Meré, J.}, {Osborne, P.}, {Pagani, C.}, {Pagano, I.}, {Pailler, F.}, {Palacin, H.}, {Palaversa, L.}, {Parsons, P.}, {Paulsen, T.}, {Pecoraro, M.}, {Pedrosa, R.}, {Pentikäinen, H.}, {Pereira, J.}, {Pichon, B.}, {Piersimoni, A. M.}, {Pineau, F.-X.}, {Plachy, E.}, {Plum, G.}, {Poujoulet, E.}, {Prsa, A.}, {Pulone, L.}, {Ragaini, S.}, {Rago, S.}, {Rambaux, N.}, {Ramos-Lerate, M.}, {Ranalli, P.}, {Rauw, G.}, {Read, A.}, {Regibo, S.}, {Renk, F.}, {Reylé, C.}, {Ribeiro, R. A.}, {Rimoldini, L.}, {Ripepi, V.}, {Riva, A.}, {Rixon, G.}, {Roelens, M.}, {Romero-Gómez, M.}, {Rowell, N.}, {Royer, F.}, {Rudolph, A.}, {Ruiz-Dern, L.}, {Sadowski, G.}, {Sagristà Sellés, T.}, {Sahlmann, J.}, {Salgado, J.}, {Salguero, E.}, {Sarasso, M.}, {Savietto, H.}, {Schnorhk, A.}, {Schultheis, M.}, {Sciacca, E.}, {Segol,
  M.}, {Segovia, J. C.}, {Segransan, D.}, {Serpell, E.}, {Shih, I-C.}, {Smareglia, R.}, {Smart, R. L.}, {Smith, C.}, {Solano, E.}, {Solitro, F.}, {Sordo, R.}, {Soria Nieto, S.}, {Souchay, J.}, {Spagna, A.}, {Spoto, F.}, {Stampa, U.}, {Steele, I. A.}, {Steidelmüller, H.}, {Stephenson, C. A.}, {Stoev, H.}, {Suess, F. F.}, {Süveges, M.}, {Surdej, J.}, {Szabados, L.}, {Szegedi-Elek, E.}, {Tapiador, D.}, {Taris, F.}, {Tauran, G.}, {Taylor, M. B.}, {Teixeira, R.}, {Terrett, D.}, {Tingley, B.}, {Trager, S. C.}, {Turon, C.}, {Ulla, A.}, {Utrilla, E.}, {Valentini, G.}, {van Elteren, A.}, {Van Hemelryck, E.}, {van Leeuwen, M.}, {Varadi, M.}, {Vecchiato, A.}, {Veljanoski, J.}, {Via, T.}, {Vicente, D.}, {Vogt, S.}, {Voss, H.}, {Votruba, V.}, {Voutsinas, S.}, {Walmsley, G.}, {Weiler, M.}, {Weingrill, K.}, {Werner, D.}, {Wevers, T.}, {Whitehead, G.}, {Wyrzykowski, L.}, {Yoldas, A.}, {Zerjal, M.}, {Zucker, S.}, {Zurbach, C.}, {Zwitter, T.}, {Alecu, A.}, {Allen, M.}, {Allende Prieto, C.}, {Amorim, A.}, {Anglada-Escudé,
  G.}, {Arsenijevic, V.}, {Azaz, S.}, {Balm, P.}, {Beck, M.}, {Bernstein, H.-H.}, {Bigot, L.}, {Bijaoui, A.}, {Blasco, C.}, {Bonfigli, M.}, {Bono, G.}, {Boudreault, S.}, {Bressan, A.}, {Brown, S.}, {Brunet, P.-M.}, {Bunclark, P.}, {Buonanno, R.}, {Butkevich, A. G.}, {Carret, C.}, {Carrion, C.}, {Chemin, L.}, {Chéreau, F.}, {Corcione, L.}, {Darmigny, E.}, {de Boer, K. S.}, {de Teodoro, P.}, {de Zeeuw, P. T.}, {Delle Luche, C.}, {Domingues, C. D.}, {Dubath, P.}, {Fodor, F.}, {Frézouls, B.}, {Fries, A.}, {Fustes, D.}, {Fyfe, D.}, {Gallardo, E.}, {Gallegos, J.}, {Gardiol, D.}, {Gebran, M.}, {Gomboc, A.}, {Gómez, A.}, {Grux, E.}, {Gueguen, A.}, {Heyrovsky, A.}, {Hoar, J.}, {Iannicola, G.}, {Isasi Parache, Y.}, {Janotto, A.-M.}, {Joliet, E.}, {Jonckheere, A.}, {Keil, R.}, {Kim, D.-W.}, {Klagyivik, P.}, {Klar, J.}, {Knude, J.}, {Kochukhov, O.}, {Kolka, I.}, {Kos, J.}, {Kutka, A.}, {Lainey, V.}, {LeBouquin, D.}, {Liu, C.}, {Loreggia, D.}, {Makarov, V. V.}, {Marseille, M. G.}, {Martayan, C.}, {Martinez-Rubi, O.},
  {Massart, B.}, {Meynadier, F.}, {Mignot, S.}, {Munari, U.}, {Nguyen, A.-T.}, {Nordlander, T.}, {Ocvirk, P.}, {O{\textbackslash}'Flaherty, K. S.}, {Olias Sanz, A.}, {Ortiz, P.}, {Osorio, J.}, {Oszkiewicz, D.}, {Ouzounis, A.}, {Palmer, M.}, {Park, P.}, {Pasquato, E.}, {Peltzer, C.}, {Peralta, J.}, {Péturaud, F.}, {Pieniluoma, T.}, {Pigozzi, E.}, {Poels, J.}, {Prat, G.}, {Prod{\textbackslash}'homme, T.}, {Raison, F.}, {Rebordao, J. M.}, {Risquez, D.}, {Rocca-Volmerange, B.}, {Rosen, S.}, {Ruiz-Fuertes, M. I.}, {Russo, F.}, {Sembay, S.}, {Serraller Vizcaino, I.}, {Short, A.}, {Siebert, A.}, {Silva, H.}, {Sinachopoulos, D.}, {Slezak, E.}, {Soffel, M.}, {Sosnowska, D.}, {Straizys, V.}, {ter Linden, M.}, {Terrell, D.}, {Theil, S.}, {Tiede, C.}, {Troisi, L.}, {Tsalmantza, P.}, {Tur, D.}, {Vaccari, M.}, {Vachier, F.}, {Valles, P.}, {Van Hamme, W.}, {Veltz, L.}, {Virtanen, J.}, {Wallut, J.-M.}, {Wichmann, R.}, {Wilkinson, M. I.}, {Ziaeepour, H.}, \& {Zschocke, S.}}]{gaia_collaboration_gaia_2016}
{Gaia Collaboration}, {Prusti, T.}, {de Bruijne, J. H. J.}, {et~al.} 2016, \bibinfo{title}{The {Gaia} mission,} A\&A, 595, A1, \dodoi{10.1051/0004-6361/201629272}

\bibitem[{ {Gaia Collaboration} {et~al.}(2018){Gaia Collaboration}, {Brown, A. G. A.}, {Vallenari, A.}, {Prusti, T.}, {de Bruijne, J. H. J.}, {Babusiaux, C.}, {Bailer-Jones, C. A. L.}, {Biermann, M.}, {Evans, D. W.}, {Eyer, L.}, {Jansen, F.}, {Jordi, C.}, {Klioner, S. A.}, {Lammers, U.}, {Lindegren, L.}, {Luri, X.}, {Mignard, F.}, {Panem, C.}, {Pourbaix, D.}, {Randich, S.}, {Sartoretti, P.}, {Siddiqui, H. I.}, {Soubiran, C.}, {van Leeuwen, F.}, {Walton, N. A.}, {Arenou, F.}, {Bastian, U.}, {Cropper, M.}, {Drimmel, R.}, {Katz, D.}, {Lattanzi, M. G.}, {Bakker, J.}, {Cacciari, C.}, {Castañeda, J.}, {Chaoul, L.}, {Cheek, N.}, {De Angeli, F.}, {Fabricius, C.}, {Guerra, R.}, {Holl, B.}, {Masana, E.}, {Messineo, R.}, {Mowlavi, N.}, {Nienartowicz, K.}, {Panuzzo, P.}, {Portell, J.}, {Riello, M.}, {Seabroke, G. M.}, {Tanga, P.}, {Thévenin, F.}, {Gracia-Abril, G.}, {Comoretto, G.}, {Garcia-Reinaldos, M.}, {Teyssier, D.}, {Altmann, M.}, {Andrae, R.}, {Audard, M.}, {Bellas-Velidis, I.}, {Benson, K.}, {Berthier, J.},
  {Blomme, R.}, {Burgess, P.}, {Busso, G.}, {Carry, B.}, {Cellino, A.}, {Clementini, G.}, {Clotet, M.}, {Creevey, O.}, {Davidson, M.}, {De Ridder, J.}, {Delchambre, L.}, {Dell{\textbackslash}'Oro, A.}, {Ducourant, C.}, {Fernández-Hernández, J.}, {Fouesneau, M.}, {Frémat, Y.}, {Galluccio, L.}, {García-Torres, M.}, {González-Núñez, J.}, {González-Vidal, J. J.}, {Gosset, E.}, {Guy, L. P.}, {Halbwachs, J.-L.}, {Hambly, N. C.}, {Harrison, D. L.}, {Hernández, J.}, {Hestroffer, D.}, {Hodgkin, S. T.}, {Hutton, A.}, {Jasniewicz, G.}, {Jean-Antoine-Piccolo, A.}, {Jordan, S.}, {Korn, A. J.}, {Krone-Martins, A.}, {Lanzafame, A. C.}, {Lebzelter, T.}, {Löffler, W.}, {Manteiga, M.}, {Marrese, P. M.}, {Martín-Fleitas, J. M.}, {Moitinho, A.}, {Mora, A.}, {Muinonen, K.}, {Osinde, J.}, {Pancino, E.}, {Pauwels, T.}, {Petit, J.-M.}, {Recio-Blanco, A.}, {Richards, P. J.}, {Rimoldini, L.}, {Robin, A. C.}, {Sarro, L. M.}, {Siopis, C.}, {Smith, M.}, {Sozzetti, A.}, {Süveges, M.}, {Torra, J.}, {van Reeven, W.}, {Abbas,
  U.}, {Abreu Aramburu, A.}, {Accart, S.}, {Aerts, C.}, {Altavilla, G.}, {Álvarez, M. A.}, {Alvarez, R.}, {Alves, J.}, {Anderson, R. I.}, {Andrei, A. H.}, {Anglada Varela, E.}, {Antiche, E.}, {Antoja, T.}, {Arcay, B.}, {Astraatmadja, T. L.}, {Bach, N.}, {Baker, S. G.}, {Balaguer-Núñez, L.}, {Balm, P.}, {Barache, C.}, {Barata, C.}, {Barbato, D.}, {Barblan, F.}, {Barklem, P. S.}, {Barrado, D.}, {Barros, M.}, {Barstow, M. A.}, {Bartholomé Muñoz, S.}, {Bassilana, J.-L.}, {Becciani, U.}, {Bellazzini, M.}, {Berihuete, A.}, {Bertone, S.}, {Bianchi, L.}, {Bienaymé, O.}, {Blanco-Cuaresma, S.}, {Boch, T.}, {Boeche, C.}, {Bombrun, A.}, {Borrachero, R.}, {Bossini, D.}, {Bouquillon, S.}, {Bourda, G.}, {Bragaglia, A.}, {Bramante, L.}, {Breddels, M. A.}, {Bressan, A.}, {Brouillet, N.}, {Brüsemeister, T.}, {Brugaletta, E.}, {Bucciarelli, B.}, {Burlacu, A.}, {Busonero, D.}, {Butkevich, A. G.}, {Buzzi, R.}, {Caffau, E.}, {Cancelliere, R.}, {Cannizzaro, G.}, {Cantat-Gaudin, T.}, {Carballo, R.}, {Carlucci, T.}, {Carrasco,
  J. M.}, {Casamiquela, L.}, {Castellani, M.}, {Castro-Ginard, A.}, {Charlot, P.}, {Chemin, L.}, {Chiavassa, A.}, {Cocozza, G.}, {Costigan, G.}, {Cowell, S.}, {Crifo, F.}, {Crosta, M.}, {Crowley, C.}, {Cuypers+, J.}, {Dafonte, C.}, {Damerdji, Y.}, {Dapergolas, A.}, {David, P.}, {David, M.}, {de Laverny, P.}, {De Luise, F.}, {De March, R.}, {de Martino, D.}, {de Souza, R.}, {de Torres, A.}, {Debosscher, J.}, {del Pozo, E.}, {Delbo, M.}, {Delgado, A.}, {Delgado, H. E.}, {Di Matteo, P.}, {Diakite, S.}, {Diener, C.}, {Distefano, E.}, {Dolding, C.}, {Drazinos, P.}, {Durán, J.}, {Edvardsson, B.}, {Enke, H.}, {Eriksson, K.}, {Esquej, P.}, {Eynard Bontemps, G.}, {Fabre, C.}, {Fabrizio, M.}, {Faigler, S.}, {Falcão, A. J.}, {Farràs Casas, M.}, {Federici, L.}, {Fedorets, G.}, {Fernique, P.}, {Figueras, F.}, {Filippi, F.}, {Findeisen, K.}, {Fonti, A.}, {Fraile, E.}, {Fraser, M.}, {Frézouls, B.}, {Gai, M.}, {Galleti, S.}, {Garabato, D.}, {García-Sedano, F.}, {Garofalo, A.}, {Garralda, N.}, {Gavel, A.}, {Gavras, P.},
  {Gerssen, J.}, {Geyer, R.}, {Giacobbe, P.}, {Gilmore, G.}, {Girona, S.}, {Giuffrida, G.}, {Glass, F.}, {Gomes, M.}, {Granvik, M.}, {Gueguen, A.}, {Guerrier, A.}, {Guiraud, J.}, {Gutiérrez-Sánchez, R.}, {Haigron, R.}, {Hatzidimitriou, D.}, {Hauser, M.}, {Haywood, M.}, {Heiter, U.}, {Helmi, A.}, {Heu, J.}, {Hilger, T.}, {Hobbs, D.}, {Hofmann, W.}, {Holland, G.}, {Huckle, H. E.}, {Hypki, A.}, {Icardi, V.}, {Janßen, K.}, {Jevardat de Fombelle, G.}, {Jonker, P. G.}, {Juhász, Á. L.}, {Julbe, F.}, {Karampelas, A.}, {Kewley, A.}, {Klar, J.}, {Kochoska, A.}, {Kohley, R.}, {Kolenberg, K.}, {Kontizas, M.}, {Kontizas, E.}, {Koposov, S. E.}, {Kordopatis, G.}, {Kostrzewa-Rutkowska, Z.}, {Koubsky, P.}, {Lambert, S.}, {Lanza, A. F.}, {Lasne, Y.}, {Lavigne, J.-B.}, {Le Fustec, Y.}, {Le Poncin-Lafitte, C.}, {Lebreton, Y.}, {Leccia, S.}, {Leclerc, N.}, {Lecoeur-Taibi, I.}, {Lenhardt, H.}, {Leroux, F.}, {Liao, S.}, {Licata, E.}, {Lindstrøm, H. E. P.}, {Lister, T. A.}, {Livanou, E.}, {Lobel, A.}, {López, M.}, {Managau,
  S.}, {Mann, R. G.}, {Mantelet, G.}, {Marchal, O.}, {Marchant, J. M.}, {Marconi, M.}, {Marinoni, S.}, {Marschalkó, G.}, {Marshall, D. J.}, {Martino, M.}, {Marton, G.}, {Mary, N.}, {Massari, D.}, {Matijevic, G.}, {Mazeh, T.}, {McMillan, P. J.}, {Messina, S.}, {Michalik, D.}, {Millar, N. R.}, {Molina, D.}, {Molinaro, R.}, {Molnár, L.}, {Montegriffo, P.}, {Mor, R.}, {Morbidelli, R.}, {Morel, T.}, {Morris, D.}, {Mulone, A. F.}, {Muraveva, T.}, {Musella, I.}, {Nelemans, G.}, {Nicastro, L.}, {Noval, L.}, {O{\textbackslash}'Mullane, W.}, {Ordénovic, C.}, {Ordóñez-Blanco, D.}, {Osborne, P.}, {Pagani, C.}, {Pagano, I.}, {Pailler, F.}, {Palacin, H.}, {Palaversa, L.}, {Panahi, A.}, {Pawlak, M.}, {Piersimoni, A. M.}, {Pineau, F.-X.}, {Plachy, E.}, {Plum, G.}, {Poggio, E.}, {Poujoulet, E.}, {Prsa, A.}, {Pulone, L.}, {Racero, E.}, {Ragaini, S.}, {Rambaux, N.}, {Ramos-Lerate, M.}, {Regibo, S.}, {Reylé, C.}, {Riclet, F.}, {Ripepi, V.}, {Riva, A.}, {Rivard, A.}, {Rixon, G.}, {Roegiers, T.}, {Roelens, M.},
  {Romero-Gómez, M.}, {Rowell, N.}, {Royer, F.}, {Ruiz-Dern, L.}, {Sadowski, G.}, {Sagristà Sellés, T.}, {Sahlmann, J.}, {Salgado, J.}, {Salguero, E.}, {Sanna, N.}, {Santana-Ros, T.}, {Sarasso, M.}, {Savietto, H.}, {Schultheis, M.}, {Sciacca, E.}, {Segol, M.}, {Segovia, J. C.}, {Ségransan, D.}, {Shih, I-C.}, {Siltala, L.}, {Silva, A. F.}, {Smart, R. L.}, {Smith, K. W.}, {Solano, E.}, {Solitro, F.}, {Sordo, R.}, {Soria Nieto, S.}, {Souchay, J.}, {Spagna, A.}, {Spoto, F.}, {Stampa, U.}, {Steele, I. A.}, {Steidelmüller, H.}, {Stephenson, C. A.}, {Stoev, H.}, {Suess, F. F.}, {Surdej, J.}, {Szabados, L.}, {Szegedi-Elek, E.}, {Tapiador, D.}, {Taris, F.}, {Tauran, G.}, {Taylor, M. B.}, {Teixeira, R.}, {Terrett, D.}, {Teyssandier, P.}, {Thuillot, W.}, {Titarenko, A.}, {Torra Clotet, F.}, {Turon, C.}, {Ulla, A.}, {Utrilla, E.}, {Uzzi, S.}, {Vaillant, M.}, {Valentini, G.}, {Valette, V.}, {van Elteren, A.}, {Van Hemelryck, E.}, {van Leeuwen, M.}, {Vaschetto, M.}, {Vecchiato, A.}, {Veljanoski, J.}, {Viala, Y.},
  {Vicente, D.}, {Vogt, S.}, {von Essen, C.}, {Voss, H.}, {Votruba, V.}, {Voutsinas, S.}, {Walmsley, G.}, {Weiler, M.}, {Wertz, O.}, {Wevers, T.}, {Wyrzykowski, L.}, {Yoldas, A.}, {Zerjal, M.}, {Ziaeepour, H.}, {Zorec, J.}, {Zschocke, S.}, {Zucker, S.}, {Zurbach, C.}, \& {Zwitter, T.}}]{gaia_collaboration_gaia_2018}
{Gaia Collaboration}, {Brown, A. G. A.}, {Vallenari, A.}, {et~al.} 2018, \bibinfo{title}{Gaia {Data} {Release} 2 - {Summary} of the contents and survey properties,} A\&A, 616, A1, \dodoi{10.1051/0004-6361/201833051}

\bibitem[{A.~E. García~Pérez {et~al.}(2016)García~Pérez, Prieto, Holtzman, Shetrone, Mészáros, Bizyaev, Carrera, Cunha, García-Hernández, Johnson, Majewski, Nidever, Schiavon, Shane, Smith, Sobeck, Troup, Zamora, Weinberg, Bovy, Eisenstein, Feuillet, Frinchaboy, Hayden, Hearty, Nguyen, O’Connell, Pinsonneault, Wilson, \& Zasowski}]{garcia_perez_aspcap_2016}
García~Pérez, A.~E., Prieto, C.~A., Holtzman, J.~A., {et~al.} 2016, \bibinfo{title}{{ASPCAP}: {THE} {APOGEE} {S}℡{LAR} {PARAMETER} {AND} {CHEMICAL} {ABUNDANCES} {PIPELINE},} The Astronomical Journal, 151, 144, \dodoi{10.3847/0004-6256/151/6/144}

\bibitem[{A. Gelman \& D.~B. Rubin(1992)Gelman \& Rubin}]{gelman_inference_1992}
Gelman, A., \& Rubin, D.~B. 1992, \bibinfo{title}{Inference from {Iterative} {Simulation} {Using} {Multiple} {Sequences},} Statistical Science, 7, 457, \dodoi{10.1214/ss/1177011136}

\bibitem[{G.~J. Gilbert {et~al.}(2022)Gilbert, MacDougall, \& Petigura}]{gilbert_implicit_2022}
Gilbert, G.~J., MacDougall, M.~G., \& Petigura, E.~A. 2022, \bibinfo{title}{Implicit {Biases} in {Transit} {Models} {Using} {Stellar} {Pseudo} {Density},} The Astronomical Journal, 164, 92, \dodoi{10.3847/1538-3881/ac7f2f}

\bibitem[{G.~J. Gilbert {et~al.}(2025)Gilbert, Petigura, \& Entrican}]{gilbert_planets_2025}
Gilbert, G.~J., Petigura, E.~A., \& Entrican, P.~M. 2025, \bibinfo{title}{Planets larger than {Neptune} have elevated eccentricities,} Proceedings of the National Academy of Sciences, 122, e2405295122, \dodoi{10.1073/pnas.2405295122}

\bibitem[{J.~E. Gunn {et~al.}(2006)Gunn, Siegmund, Mannery, Owen, Hull, Leger, Carey, Knapp, York, Boroski, Kent, Lupton, Rockosi, Evans, Waddell, Anderson, Annis, Barentine, Bartoszek, Bastian, Bracker, Brewington, Briegel, Brinkmann, Brown, Carr, Czarapata, Drennan, Dombeck, Federwitz, Gillespie, Gonzales, Hansen, Harvanek, Hayes, Jordan, Kinney, Klaene, Kleinman, Kron, Kresinski, Lee, Limmongkol, Lindenmeyer, Long, Loomis, McGehee, Mantsch, Eric H.~Neilsen, Neswold, Newman, Nitta, John~Peoples, Pier, Prieto, Prosapio, Rivetta, Schneider, Snedden, \& Wang}]{gunn_25_2006}
Gunn, J.~E., Siegmund, W.~A., Mannery, E.~J., {et~al.} 2006, \bibinfo{title}{The 2.5 m {Telescope} of the {Sloan} {Digital} {Sky} {Survey},} The Astronomical Journal, 131, 2332, \dodoi{10.1086/500975}

\bibitem[{C.~R. Harris {et~al.}(2020)Harris, Millman, van~der Walt, Gommers, Virtanen, Cournapeau, Wieser, Taylor, Berg, Smith, Kern, Picus, Hoyer, van Kerkwijk, Brett, Haldane, del Río, Wiebe, Peterson, Gérard-Marchant, Sheppard, Reddy, Weckesser, Abbasi, Gohlke, \& Oliphant}]{harris_array_2020}
Harris, C.~R., Millman, K.~J., van~der Walt, S.~J., {et~al.} 2020, \bibinfo{title}{Array programming with {NumPy},} Nature, 585, 357, \dodoi{10.1038/s41586-020-2649-2}

\bibitem[{S. Hattori {et~al.}(2024)Hattori, Garcia, Murray, Dong, Dholakia, Degen, \& Foreman-Mackey}]{hattori_exoplanet-devjaxoplanet_2024}
Hattori, S., Garcia, L., Murray, C., {et~al.} 2024, exoplanet-dev/jaxoplanet, exoplanet.
\newblock \url{https://github.com/exoplanet-dev/jaxoplanet}

\bibitem[{M.~Y. He {et~al.}(2020)He, Ford, Ragozzine, \& Carrera}]{he_architectures_2020}
He, M.~Y., Ford, E.~B., Ragozzine, D., \& Carrera, D. 2020, \bibinfo{title}{Architectures of {Exoplanetary} {Systems}. {III}. {Eccentricity} and {Mutual} {Inclination} {Distributions} of {AMD}-stable {Planetary} {Systems},} The Astronomical Journal, 160, 276, \dodoi{10.3847/1538-3881/abba18}

\bibitem[{E. Higson {et~al.}(2019)Higson, Handley, Hobson, \& Lasenby}]{higson_dynamic_2019}
Higson, E., Handley, W., Hobson, M., \& Lasenby, A. 2019, \bibinfo{title}{Dynamic nested sampling: an improved algorithm for parameter estimation and evidence calculation,} Statistics and Computing, 29, 891, \dodoi{10.1007/s11222-018-9844-0}

\bibitem[{T. Holczer {et~al.}(2016)Holczer, Mazeh, Nachmani, Jontof-Hutter, Ford, Fabrycky, Ragozzine, Kane, \& Steffen}]{holczer_transit_2016}
Holczer, T., Mazeh, T., Nachmani, G., {et~al.} 2016, \bibinfo{title}{{TRANSIT} {TIMING} {OBSERVATIONS} {FROM} \textit{{KEPLER}} . {IX}. {CATALOG} {OF} {THE} {FULL} {LONG}-{CADENCE} {DATA} {SET},} The Astrophysical Journal Supplement Series, 225, 9, \dodoi{10.3847/0067-0049/225/1/9}

\bibitem[{T.-O. Husser {et~al.}(2013)Husser, Berg, Dreizler, Homeier, Reiners, Barman, \& Hauschildt}]{husser_new_2013}
Husser, T.-O., Berg, S. W.-v., Dreizler, S., {et~al.} 2013, \bibinfo{title}{A new extensive library of {PHOENIX} stellar atmospheres and synthetic spectra,} Astronomy \& Astrophysics, 553, A6, \dodoi{10.1051/0004-6361/201219058}

\bibitem[{S.~A. Johnson {et~al.}(2020)Johnson, Penny, Gaudi, Kerins, Rattenbury, Robin, Calchi~Novati, \& Henderson}]{johnson_predictions_2020}
Johnson, S.~A., Penny, M., Gaudi, B.~S., {et~al.} 2020, \bibinfo{title}{Predictions of the {Nancy} {Grace} {Roman} {Space} {Telescope} {Galactic} {Exoplanet} {Survey}. {II}. {Free}-floating {Planet} {Detection} {Rates}*,} The Astronomical Journal, 160, 123, \dodoi{10.3847/1538-3881/aba75b}

\bibitem[{N.~A. Kaib {et~al.}(2013)Kaib, Raymond, \& Duncan}]{kaib_planetary_2013}
Kaib, N.~A., Raymond, S.~N., \& Duncan, M. 2013, \bibinfo{title}{Planetary {System} {Disruption} by {Galactic} {Perturbations} to {Wide} {Binary} {Stars},} Nature, 493, 381, \dodoi{10.1038/nature11780}

\bibitem[{D.~M. Kipping(2014)Kipping}]{kipping_bayesian_2014}
Kipping, D.~M. 2014, \bibinfo{title}{Bayesian priors for the eccentricity of transiting planets,} Monthly Notices of the Royal Astronomical Society, 444, 2263, \dodoi{10.1093/mnras/stu1561}

\bibitem[{D.~M. Kipping {et~al.}(2012)Kipping, Dunn, Jasinski, \& Manthri}]{kipping_novel_2012}
Kipping, D.~M., Dunn, W.~R., Jasinski, J.~M., \& Manthri, V.~P. 2012, \bibinfo{title}{A novel method to photometrically constrain orbital eccentricities: {Multibody} {Asterodensity} {Profiling},} Monthly Notices of the Royal Astronomical Society, 421, 1166, \dodoi{10.1111/j.1365-2966.2011.20376.x}

\bibitem[{D.~G. Koch {et~al.}(2010)Koch, Borucki, Basri, Batalha, Brown, Caldwell, Christensen-Dalsgaard, Cochran, DeVore, Dunham, Gautier, Geary, Gilliland, Gould, Jenkins, Kondo, Latham, Lissauer, Marcy, Monet, Sasselov, Boss, Brownlee, Caldwell, Dupree, Howell, Kjeldsen, Meibom, Morrison, Owen, Reitsema, Tarter, Bryson, Dotson, Gazis, Haas, Kolodziejczak, Rowe, Van~Cleve, Allen, Chandrasekaran, Clarke, Li, Quintana, Tenenbaum, Twicken, \& Wu}]{koch_kepler_2010}
Koch, D.~G., Borucki, W.~J., Basri, G., {et~al.} 2010, \bibinfo{title}{{KEPLER} {MISSION} {DESIGN}, {REALIZED} {PHOTOMETRIC} {PERFORMANCE}, {AND} {EARLY} {SCIENCE},} The Astrophysical Journal Letters, 713, L79, \dodoi{10.1088/2041-8205/713/2/L79}

\bibitem[{S. Koposov {et~al.}(2024)Koposov, Speagle, Barbary, Ashton, Bennett, Buchner, Scheffler, Cook, Talbot, Guillochon, Cubillos, Ramos, Dartiailh, Ilya, Tollerud, Lang, Johnson, jtmendel, Higson, Vandal, Daylan, Angus, patelR, Cargile, Sheehan, Pitkin, Kirk, Leja, joezuntz, \& Goldstein}]{koposov_joshspeagledynesty_2024}
Koposov, S., Speagle, J., Barbary, K., {et~al.} 2024, joshspeagle/dynesty: v2.1.4, Zenodo, \dodoi{10.5281/zenodo.12537467}

\bibitem[{J.~M.~D. Kruijssen {et~al.}(2020)Kruijssen, Longmore, \& Chevance}]{kruijssen_bridging_2020}
Kruijssen, J. M.~D., Longmore, S.~N., \& Chevance, M. 2020, \bibinfo{title}{Bridging the {Planet} {Radius} {Valley}: {Stellar} {Clustering} as a {Key} {Driver} for {Turning} {Sub}-{Neptunes} into {Super}-{Earths},} The Astrophysical Journal Letters, 905, L18, \dodoi{10.3847/2041-8213/abccc3}

\bibitem[{J.~M.~D. Kruijssen {et~al.}(2021)Kruijssen, Longmore, Chevance, Laporte, Motylinski, Keller, \& Henshaw}]{kruijssen_not_2021}
Kruijssen, J. M.~D., Longmore, S.~N., Chevance, M., {et~al.} 2021, Not the {Birth} {Cluster}: the {Stellar} {Clustering} that {Shapes} {Planetary} {Systems} is {Generated} by {Galactic}-{Dynamical} {Perturbations}, arXiv, \dodoi{10.48550/arXiv.2109.06182}

\bibitem[{M.~A. Limbach \& E.~L. Turner(2015)Limbach \& Turner}]{limbach_exoplanet_2015}
Limbach, M.~A., \& Turner, E.~L. 2015, \bibinfo{title}{Exoplanet orbital eccentricity: {Multiplicity} relation and the {Solar} {System},} Proceedings of the National Academy of Sciences, 112, 20, \dodoi{10.1073/pnas.1406545111}

\bibitem[{M.~G. MacDougall {et~al.}(2023)MacDougall, Gilbert, \& Petigura}]{macdougall_accurate_2023}
MacDougall, M.~G., Gilbert, G.~J., \& Petigura, E.~A. 2023, \bibinfo{title}{Accurate and {Efficient} {Photoeccentric} {Transit} {Modeling},} The Astronomical Journal, 166, 61, \dodoi{10.3847/1538-3881/ace16d}

\bibitem[{S.~R. Majewski {et~al.}(2017)Majewski, Schiavon, Frinchaboy, Prieto, Barkhouser, Bizyaev, Blank, Brunner, Burton, Carrera, Chojnowski, Cunha, Epstein, Fitzgerald, Pérez, Hearty, Henderson, Holtzman, Johnson, Lam, Lawler, Maseman, Mészáros, Nelson, Nguyen, Nidever, Pinsonneault, Shetrone, Smee, Smith, Stolberg, Skrutskie, Walker, Wilson, Zasowski, Anders, Basu, Beland, Blanton, Bovy, Brownstein, Carlberg, Chaplin, Chiappini, Eisenstein, Elsworth, Feuillet, Fleming, Galbraith-Frew, García, García-Hernández, Gillespie, Girardi, Gunn, Hasselquist, Hayden, Hekker, Ivans, Kinemuchi, Klaene, Mahadevan, Mathur, Mosser, Muna, Munn, Nichol, O’Connell, Parejko, Robin, Rocha-Pinto, Schultheis, Serenelli, Shane, Aguirre, Sobeck, Thompson, Troup, Weinberg, \& Zamora}]{majewski_apache_2017}
Majewski, S.~R., Schiavon, R.~P., Frinchaboy, P.~M., {et~al.} 2017, \bibinfo{title}{The {Apache} {Point} {Observatory} {Galactic} {Evolution} {Experiment} ({APOGEE}),} The Astronomical Journal, 154, 94, \dodoi{10.3847/1538-3881/aa784d}

\bibitem[{T. Mazeh {et~al.}(2013)Mazeh, Nachmani, Holczer, Fabrycky, Ford, Sanchis-Ojeda, Sokol, Rowe, Zucker, Agol, Carter, Lissauer, Quintana, Ragozzine, Steffen, \& Welsh}]{mazeh_transit_2013}
Mazeh, T., Nachmani, G., Holczer, T., {et~al.} 2013, \bibinfo{title}{{TRANSIT} {TIMING} {OBSERVATIONS} {FROM} \textit{{KEPLER}} . {VIII}. {CATALOG} {OF} {TRANSIT} {TIMING} {MEASUREMENTS} {OF} {THE} {FIRST} {TWELVE} {QUARTERS},} The Astrophysical Journal Supplement Series, 208, 16, \dodoi{10.1088/0067-0049/208/2/16}

\bibitem[{M.~A.~S. McTier {et~al.}(2020)McTier, Kipping, \& Johnston}]{mctier_8_2020}
McTier, M. A.~S., Kipping, D.~M., \& Johnston, K. 2020, \bibinfo{title}{8 in 10 {Stars} in the {Milky} {Way} {Bulge} experience stellar encounters within 1000 {AU} in a gigayear,} Monthly Notices of the Royal Astronomical Society, 495, 2105, \dodoi{10.1093/mnras/staa1232}

\bibitem[{S.~C. Millholland {et~al.}(2021)Millholland, He, Ford, Ragozzine, Fabrycky, \& Winn}]{millholland_evidence_2021}
Millholland, S.~C., He, M.~Y., Ford, E.~B., {et~al.} 2021, \bibinfo{title}{Evidence for a {Nondichotomous} {Solution} to the {Kepler} {Dichotomy}: {Mutual} {Inclinations} of {Kepler} {Planetary} {Systems} from {Transit} {Duration} {Variations},} The Astronomical Journal, 162, 166, \dodoi{10.3847/1538-3881/ac0f7a}

\bibitem[{S.~M. Mills {et~al.}(2019)Mills, Howard, Petigura, Fulton, Isaacson, \& Weiss}]{mills_california-kepler_2019}
Mills, S.~M., Howard, A.~W., Petigura, E.~A., {et~al.} 2019, \bibinfo{title}{The {California}-{Kepler} {Survey}. {VIII}. {Eccentricities} of \textit{{Kepler}} {Planets} and {Tentative} {Evidence} of a {High}-metallicity {Preference} for {Small} {Eccentric} {Planets},} The Astronomical Journal, 157, 198, \dodoi{10.3847/1538-3881/ab1009}

\bibitem[{V. Nagpal {et~al.}(2023)Nagpal, Blunt, Bowler, Dupuy, Nielsen, \& Wang}]{nagpal_impact_2023}
Nagpal, V., Blunt, S., Bowler, B.~P., {et~al.} 2023, \bibinfo{title}{The {Impact} of {Bayesian} {Hyperpriors} on the {Population}-{Level} {Eccentricity} {Distribution} of {Imaged} {Planets},} The Astronomical Journal, 165, 32, \dodoi{10.3847/1538-3881/ac9fd2}

\bibitem[{ {NASA Exoplanet Archive}(2019){NASA Exoplanet Archive}}]{nasa_exoplanet_archive_kepler_2019}
{NASA Exoplanet Archive}. 2019, Kepler {Objects} of {Interest} {Cumulative} {Table}, IPAC, \dodoi{10.26133/NEA4}

\bibitem[{D.~L. Nidever {et~al.}(2015)Nidever, Holtzman, Prieto, Beland, Bender, Bizyaev, Burton, Desphande, Fleming, García~Pérez, Hearty, Majewski, Mészáros, Muna, Nguyen, Schiavon, Shetrone, Skrutskie, Sobeck, \& Wilson}]{nidever_data_2015}
Nidever, D.~L., Holtzman, J.~A., Prieto, C.~A., {et~al.} 2015, \bibinfo{title}{{THE} {DATA} {REDUCTION} {PIPELINE} {FOR} {THE} {APACHE} {POINT} {OBSERVATORY} {GALACTIC} {EVOLUTION} {EXPERIMENT},} The Astronomical Journal, 150, 173, \dodoi{10.1088/0004-6256/150/6/173}

\bibitem[{J. Nielsen {et~al.}(2023)Nielsen, Gent, Bergemann, Eitner, \& Johansen}]{nielsen_planet_2023}
Nielsen, J., Gent, M.~R., Bergemann, M., Eitner, P., \& Johansen, A. 2023, \bibinfo{title}{Planet formation throughout the {Milky} {Way} - {Planet} populations in the context of {Galactic} chemical evolution,} Astronomy \& Astrophysics, 678, A74, \dodoi{10.1051/0004-6361/202346697}

\bibitem[{H. Parviainen \& S. Aigrain(2015)Parviainen \& Aigrain}]{parviainen_ldtk_2015}
Parviainen, H., \& Aigrain, S. 2015, \bibinfo{title}{ldtk: {Limb} {Darkening} {Toolkit},} Monthly Notices of the Royal Astronomical Society, 453, 3821, \dodoi{10.1093/mnras/stv1857}

\bibitem[{D. Phan {et~al.}(2019)Phan, Pradhan, \& Jankowiak}]{phan_composable_2019}
Phan, D., Pradhan, N., \& Jankowiak, M. 2019, Composable {Effects} for {Flexible} and {Accelerated} {Probabilistic} {Programming} in {NumPyro}, \dodoi{10.48550/arXiv.1912.11554}

\bibitem[{B. Pu~(濮勃南) \& Y. Wu~(武延庆)(2015)Pu~(濮勃南) \& Wu~(武延庆)}]{pu__spacing_2015}
Pu~(濮勃南), B., \& Wu~(武延庆), Y. 2015, \bibinfo{title}{{SPACING} {OF} {KEPLER} {PLANETS}: {SCULPTING} {BY} {DYNAMICAL} {INSTABILITY},} The Astrophysical Journal, 807, 44, \dodoi{10.1088/0004-637X/807/1/44}

\bibitem[{H. Rauer {et~al.}(2025)Rauer, Aerts, Cabrera, Deleuil, Erikson, Gizon, Goupil, Heras, Walloschek, Lorenzo-Alvarez, Marliani, Martin-Garcia, Mas-Hesse, O’Rourke, Osborn, Pagano, Piotto, Pollacco, Ragazzoni, Ramsay, Udry, Appourchaux, Benz, Brandeker, Güdel, Janot-Pacheco, Kabath, Kjeldsen, Min, Santos, Smith, Suarez, Werner, Aboudan, Abreu, Acuña, Adams, Adibekyan, Affer, Agneray, Agnor, Aguirre Børsen-Koch, Ahmed, Aigrain, Al-Bahlawan, Alcacera~Gil, Alei, Alencar, Alexander, Alfonso-Garzón, Alibert, Allende~Prieto, Almeida, Alonso~Sobrino, Altavilla, Althaus, Alvarez~Trujillo, Amarsi, Ammler-von Eiff, Amôres, Andrade, Antoniadis-Karnavas, António, Aparicio~del Moral, Appolloni, Arena, Armstrong, Aroca~Aliaga, Asplund, Audenaert, Auricchio, Avelino, Baeke, Baillié, Balado, Ballber~Balagueró, Balestra, Ball, Ballans, Ballot, Barban, Barbary, Barbieri, Barceló~Forteza, Barker, Barklem, Barnes, Barrado~Navascues, Barragan, Baruteau, Basu, Baudin, Baumeister, Bayliss, Bazot, Beck, Belkacem,
  Bellinger, Benatti, Benomar, Bérard, Bergemann, Bergomi, Bernardo, Biazzo, Bignamini, Bigot, Billot, Binet, Biondi, Biondi, Birch, Bitsch, Bluhm~Ceballos, Bódi, Bognár, Boisse, Bolmont, Bonanno, Bonavita, Bonfanti, Bonfils, Bonito, Bonomo, Börner, Boro~Saikia, Borreguero~Martín, Borsa, Borsato, Bossini, Bouchy, Boué, Boufleur, Boumier, Bourrier, Bowman, Bozzo, Bradley, Bray, Bressan, Breton, Brienza, Brito, Brogi, Brown, Brown, Brun, Bruno, Bruns, Buchhave, Bugnet, Buldgen, Burgess, Busatta, Busso, Buzasi, Caballero, Cabral, Cabrero~Gomez, Calderone, Cameron, Cameron, Campante, Campos~Gestal, Canto~Martins, Cara, Carone, Carrasco, Casagrande, Casewell, Cassisi, Castellani, Castro, Catala, Catalán~Fernández, Catelan, Cegla, Cerruti, Cessa, Chadid, Chaplin, Charpinet, Chiappini, Chiarucci, Chiavassa, Chinellato, Chirulli, Christensen-Dalsgaard, Church, Claret, Clarke, Claudi, Clermont, Coelho, Coelho, Cogato, Colomé, Condamin, Conde~García, Conseil, Corbard, Correia, Corsaro, Cosentino, Costes,
  Cottinelli, Covone, Creevey, Crida, Csizmadia, Cunha, Curry, da~Costa, da~Silva, Dalal, Damasso, Damiani, Damiani, das Chagas, Davies, Davies, Davies, Davison, de~Almeida, de~Angeli, de~Barros, de~CastroLeão, de~Freitas, de~Freitas, De~Martino, de~Medeiros, de~Paula, de~Pedraza~Gómez, de~Plaa, De~Ridder, Deal, Decin, Deeg, Degl’Innocenti, Deheuvels, del Burgo, Del~Sordo, Delgado-Mena, Demangeon, Denk, Derekas, Desert, Desidera, Dexet, Di~Criscienzo, Di~Giorgio, Di~Mauro, Diaz~Rial, Díaz-García, Dima, Dinuzzi, Dionatos, Distefano, do~Nascimento~Jr., Domingo, D’Orazi, Dorn, Doyle, Duarte, Ducellier, Dumaye, Dumusque, Dupret, Eggenberger, Ehrenreich, Eigmüller, Eising, Emilio, Eriksson, Ermocida, Escate~Giribaldi, Eschen, Espinosa~Yáñez, Estrela, Evans, Fabbian, Fabrizio, Faria, Farina, Farinato, Feliz, Feltzing, Fenouillet, Fernández, Ferrari, Ferraz-Mello, Fialho, Fienga, Figueira, Fiori, Flaccomio, Focardi, Foley, Fontignie, \& Ford}]{rauer_plato_2025}
Rauer, H., Aerts, C., Cabrera, J., {et~al.} 2025, \bibinfo{title}{The {PLATO} mission,} Experimental Astronomy, 59, 26, \dodoi{10.1007/s10686-025-09985-9}

\bibitem[{T.~P. Robitaille {et~al.}(2013)Robitaille, Tollerud, Greenfield, Droettboom, Bray, Aldcroft, Davis, Ginsburg, Price-Whelan, Kerzendorf, Conley, Crighton, Barbary, Muna, Ferguson, Grollier, Parikh, Nair, Günther, Deil, Woillez, Conseil, Kramer, Turner, Singer, Fox, Weaver, Zabalza, Edwards, Bostroem, Burke, Casey, Crawford, Dencheva, Ely, Jenness, Labrie, Lim, Pierfederici, Pontzen, Ptak, Refsdal, Servillat, \& Streicher}]{robitaille_astropy_2013}
Robitaille, T.~P., Tollerud, E.~J., Greenfield, P., {et~al.} 2013, \bibinfo{title}{Astropy: {A} community {Python} package for astronomy,} Astronomy \& Astrophysics, 558, A33, \dodoi{10.1051/0004-6361/201322068}

\bibitem[{L. Rodet {et~al.}(2021)Rodet, Su, \& Lai}]{rodet_correlation_2021}
Rodet, L., Su, Y., \& Lai, D. 2021, \bibinfo{title}{On the {Correlation} between {Hot} {Jupiters} and {Stellar} {Clustering}: {High}-eccentricity {Migration} {Induced} by {Stellar} {Flybys},} The Astrophysical Journal, 913, 104, \dodoi{10.3847/1538-4357/abf8a7}

\bibitem[{S. Sagear \& S. Ballard(2023)Sagear \& Ballard}]{sagear_orbital_2023}
Sagear, S., \& Ballard, S. 2023, \bibinfo{title}{The orbital eccentricity distribution of planets orbiting {M} dwarfs,} Proceedings of the National Academy of Sciences, 120, e2217398120, \dodoi{10.1073/pnas.2217398120}

\bibitem[{S. Sagear {et~al.}(2025)Sagear, Ballard, Gilbert, Albornoz, \& Lam}]{sagear_orbital_2025}
Sagear, S., Ballard, S., Gilbert, G.~J., Albornoz, M., \& Lam, C. 2025, The {Orbital} {Eccentricity}-{Radius} {Relation} for {Planets} {Orbiting} {M} {Dwarfs}, arXiv, \dodoi{10.48550/arXiv.2507.07169}

\bibitem[{F.~A. Santana {et~al.}(2021)Santana, Beaton, Covey, O’Connell, Longa-Peña, Cohen, Fernández-Trincado, Hayes, Zasowski, Sobeck, Majewski, Chojnowski, De~Lee, Oelkers, Stringfellow, Almeida, Anguiano, Donor, Frinchaboy, Hasselquist, Johnson, Kollmeier, Nidever, Price-Whelan, Rojas-Arriagada, Schultheis, Shetrone, Simon, Aerts, Borissova, Drout, Geisler, Law, Medina, Minniti, Monachesi, Muñoz, Poleski, Roman-Lopes, Schlaufman, Stutz, Teske, Tkachenko, Van~Saders, Weinberger, \& Zoccali}]{santana_final_2021}
Santana, F.~A., Beaton, R.~L., Covey, K.~R., {et~al.} 2021, \bibinfo{title}{Final {Targeting} {Strategy} for the {SDSS}-{IV} {APOGEE}-{2S} {Survey},} The Astronomical Journal, 162, 303, \dodoi{10.3847/1538-3881/ac2cbc}

\bibitem[{C. Schoettler \& J.~E. Owen(2024)Schoettler \& Owen}]{schoettler_effect_2024}
Schoettler, C., \& Owen, J.~E. 2024, \bibinfo{title}{The effect of dynamical interactions in stellar birth environments on the orbits of young close-in planetary systems,} Monthly Notices of the Royal Astronomical Society, 533, 3484, \dodoi{10.1093/mnras/stae1900}

\bibitem[{G. Schwarz(1978)Schwarz}]{schwarz_estimating_1978}
Schwarz, G. 1978, \bibinfo{title}{Estimating the {Dimension} of a {Model},} The Annals of Statistics, 6, \dodoi{10.1214/aos/1176344136}

\bibitem[{M. Shetrone {et~al.}(2015)Shetrone, Bizyaev, Lawler, Prieto, Johnson, Smith, Cunha, Holtzman, Pérez, Mészáros, Sobeck, Zamora, García-Hernández, Souto, Chojnowski, Koesterke, Majewski, \& Zasowski}]{shetrone_sdss-iii_2015}
Shetrone, M., Bizyaev, D., Lawler, J.~E., {et~al.} 2015, \bibinfo{title}{{THE} {SDSS}-{III} {APOGEE} {SPECTRAL} {LINE} {LIST} {FOR} {H}-{BAND} {SPECTROSCOPY},} The Astrophysical Journal Supplement Series, 221, 24, \dodoi{10.1088/0067-0049/221/2/24}

\bibitem[{J. Skilling(2004)Skilling}]{skilling_nested_2004}
Skilling, J. 2004, \bibinfo{title}{Nested {Sampling},} AIP Conference Proceedings, 735, 395, \dodoi{10.1063/1.1835238}

\bibitem[{J. Skilling(2006)Skilling}]{skilling_nested_2006}
Skilling, J. 2006, \bibinfo{title}{Nested sampling for general {Bayesian} computation,} Bayesian Analysis, 1, 833, \dodoi{10.1214/06-BA127}

\bibitem[{V.~V. Smith {et~al.}(2021)Smith, Bizyaev, Cunha, Shetrone, Souto, Allende~Prieto, Masseron, Mészáros, Jönsson, Hasselquist, Osorio, García-Hernández, Plez, Beaton, Holtzman, Majewski, Stringfellow, \& Sobeck}]{smith_apogee_2021}
Smith, V.~V., Bizyaev, D., Cunha, K., {et~al.} 2021, \bibinfo{title}{The {APOGEE} {Data} {Release} 16 {Spectral} {Line} {List},} The Astronomical Journal, 161, 254, \dodoi{10.3847/1538-3881/abefdc}

\bibitem[{J.~S. Speagle(2020)Speagle}]{speagle_dynesty_2020}
Speagle, J.~S. 2020, \bibinfo{title}{dynesty: a dynamic nested sampling package for estimating {Bayesian} posteriors and evidences,} Monthly Notices of the Royal Astronomical Society, 493, 3132, \dodoi{10.1093/mnras/staa278}

\bibitem[{V. Van~Eylen {et~al.}(2019)Van~Eylen, Albrecht, Huang, MacDonald, Dawson, Cai, Foreman-Mackey, Lundkvist, Aguirre, Snellen, \& Winn}]{van_eylen_orbital_2019}
Van~Eylen, V., Albrecht, S., Huang, X., {et~al.} 2019, \bibinfo{title}{The {Orbital} {Eccentricity} of {Small} {Planet} {Systems},} The Astronomical Journal, 157, 61, \dodoi{10.3847/1538-3881/aaf22f}

\bibitem[{D. Veras \& N.~W. Evans(2013)Veras \& Evans}]{veras_exoplanets_2013}
Veras, D., \& Evans, N.~W. 2013, \bibinfo{title}{Exoplanets beyond the {Solar} neighbourhood: {Galactic} tidal perturbations,} Monthly Notices of the Royal Astronomical Society, 430, 403, \dodoi{10.1093/mnras/sts647}

\bibitem[{J.~C. Wilson {et~al.}(2019)Wilson, Hearty, Skrutskie, Majewski, Holtzman, Eisenstein, Gunn, Blank, Henderson, Smee, Nelson, Nidever, Arns, Barkhouser, Barr, Beland, Bershady, Blanton, Brunner, Burton, Carey, Carr, Colque, Crane, Damke, Davidson, Dean, Di~Mille, Don, Ebelke, Evans, Fitzgerald, Gillespie, Hall, Harding, Harding, Hammond, Hancock, Harrison, Hope, Horne, Karakla, Lam, Leger, MacDonald, Maseman, Matsunari, Melton, Mitcheltree, O’Brien, O’Connell, Patten, Richardson, Rieke, Rieke, Roman-Lopes, Schiavon, Sobeck, Stolberg, Stoll, Tembe, Trujillo, Uomoto, Vernieri, Walker, Weinberg, Young, Anthony-Brumfield, Bizyaev, Breslauer, Lee, Downey, Halverson, Huehnerhoff, Klaene, Leon, Long, Mahadevan, Malanushenko, Nguyen, Owen, Sánchez-Gallego, Sayres, Shane, Shectman, Shetrone, Skinner, Stauffer, \& Zhao}]{wilson_apache_2019}
Wilson, J.~C., Hearty, F.~R., Skrutskie, M.~F., {et~al.} 2019, \bibinfo{title}{The {Apache} {Point} {Observatory} {Galactic} {Evolution} {Experiment} ({APOGEE}) {Spectrographs},} Publications of the Astronomical Society of the Pacific, 131, 055001, \dodoi{10.1088/1538-3873/ab0075}

\bibitem[{R.~F. Wilson {et~al.}(2023)Wilson, Barclay, Powell, Schlieder, Hedges, Montet, Quintana, Mcdonald, Penny, Espinoza, \& Kerins}]{wilson_transiting_2023}
Wilson, R.~F., Barclay, T., Powell, B.~P., {et~al.} 2023, \bibinfo{title}{Transiting {Exoplanet} {Yields} for the {Roman} {Galactic} {Bulge} {Time} {Domain} {Survey} {Predicted} from {Pixel}-level {Simulations},} The Astrophysical Journal Supplement Series, 269, 5, \dodoi{10.3847/1538-4365/acf3df}

\bibitem[{A.~J. Winter {et~al.}(2020)Winter, Kruijssen, Longmore, \& Chevance}]{winter_stellar_2020}
Winter, A.~J., Kruijssen, J. M.~D., Longmore, S.~N., \& Chevance, M. 2020, \bibinfo{title}{Stellar clustering shapes the architecture of planetary systems,} Nature, 586, 528, \dodoi{10.1038/s41586-020-2800-0}

\bibitem[{J.-W. Xie {et~al.}(2016)Xie, Dong, Zhu, Huber, Zheng, De~Cat, Fu, Liu, Luo, Wu, Zhang, Zhang, Zhou, Cao, Hou, Wang, \& Zhang}]{xie_exoplanet_2016}
Xie, J.-W., Dong, S., Zhu, Z., {et~al.} 2016, \bibinfo{title}{Exoplanet orbital eccentricities derived from {LAMOST}–{Kepler} analysis,} Proceedings of the National Academy of Sciences, 113, 11431, \dodoi{10.1073/pnas.1604692113}

\bibitem[{J.-Y. Yang {et~al.}(2023)Yang, Chen, Xie, Zhou, Dong, Zhu, Zheng, Liu, Zong, \& Luo}]{yang_planets_2023}
Yang, J.-Y., Chen, D.-C., Xie, J.-W., {et~al.} 2023, \bibinfo{title}{Planets {Across} {Space} and {Time} ({PAST}). {IV}. {The} {Occurrence} and {Architecture} of {Kepler} {Planetary} {Systems} as a {Function} of {Kinematic} {Age} {Revealed} by the {LAMOST}–{Gaia}–{Kepler} {Sample},} The Astronomical Journal, 166, 243, \dodoi{10.3847/1538-3881/ad0368}

\bibitem[{N.~L. Zakamska \& S. Tremaine(2004)Zakamska \& Tremaine}]{zakamska_excitation_2004}
Zakamska, N.~L., \& Tremaine, S. 2004, \bibinfo{title}{Excitation and {Propagation} of {Eccentricity} {Disturbances} in {Planetary} {Systems},} The Astronomical Journal, 128, 869, \dodoi{10.1086/422023}

\bibitem[{G. Zasowski {et~al.}(2013)Zasowski, Johnson, Frinchaboy, Majewski, Nidever, Pinto, Girardi, Andrews, Chojnowski, Cudworth, Jackson, Munn, Skrutskie, Beaton, Blake, Covey, Deshpande, Epstein, Fabbian, Fleming, Hernandez, Herrero, Mahadevan, Mészáros, Schultheis, Sellgren, Terrien, van Saders, Prieto, Bizyaev, Burton, Cunha, da~Costa, Hasselquist, Hearty, Holtzman, García~Pérez, Maia, O'Connell, O'Donnell, Pinsonneault, Santiago, Schiavon, Shetrone, Smith, \& Wilson}]{zasowski_target_2013}
Zasowski, G., Johnson, J.~A., Frinchaboy, P.~M., {et~al.} 2013, \bibinfo{title}{{TARGET} {SELECTION} {FOR} {THE} {APACHE} {POINT} {OBSERVATORY} {GALACTIC} {EVOLUTION} {EXPERIMENT} ({APOGEE}),} The Astronomical Journal, 146, 81, \dodoi{10.1088/0004-6256/146/4/81}

\bibitem[{G. Zasowski {et~al.}(2017)Zasowski, Cohen, Chojnowski, Santana, Oelkers, Andrews, Beaton, Bender, Bird, Bovy, Carlberg, Covey, Cunha, Dell’Agli, Fleming, Frinchaboy, García-Hernández, Harding, Holtzman, Johnson, Kollmeier, Majewski, Mészáros, Munn, Muñoz, Ness, Nidever, Poleski, Román-Zúñiga, Shetrone, Simon, Smith, Sobeck, Stringfellow, Szigetiáros, Tayar, \& Troup}]{zasowski_target_2017}
Zasowski, G., Cohen, R.~E., Chojnowski, S.~D., {et~al.} 2017, \bibinfo{title}{Target {Selection} for the {SDSS}-{IV} {APOGEE}-2 {Survey},} The Astronomical Journal, 154, 198, \dodoi{10.3847/1538-3881/aa8df9}

\bibitem[{J.~K. Zink {et~al.}(2023)Zink, Hardegree-Ullman, Christiansen, Petigura, Boley, Bhure, Rice, Yee, Isaacson, Fernandes, Howard, Blunt, Lubin, Chontos, Pidhorodetska, \& MacDougall}]{zink_scaling_2023}
Zink, J.~K., Hardegree-Ullman, K.~K., Christiansen, J.~L., {et~al.} 2023, \bibinfo{title}{Scaling {K2}. {VI}. {Reduced} {Small} {Planet} {Occurrence} in {High} {Galactic} {Amplitude} {Stars},} The Astronomical Journal, 165, 262, \dodoi{10.3847/1538-3881/acd24c}

\end{thebibliography}
\bibliographystyle{aasjournalv7}

\end{CJK}
\end{document}